\documentclass[amsmath,amssymb,aps,pra,twocolumn,superscriptaddress,nofootinbib,longbibliography]{revtex4-2}

\usepackage{graphicx}
\usepackage{subcaption}
\usepackage{dcolumn}
\usepackage{bm}
\usepackage{siunitx}
\usepackage[colorlinks=true,linkcolor=black,citecolor=blue,urlcolor=black]{hyperref}
\usepackage{comment}
\usepackage{mathtools}
\usepackage{bbold}
\usepackage{ragged2e}
\usepackage{float} 

\DeclarePairedDelimiter\norm{\lvert}{\rvert}      

\DeclareSIUnit\atomicmassunit{u}
\DeclareSIUnit\bar{bar}

\usepackage{color}
\usepackage{xcolor}

\begin{document}


\title{Shell formation and two-dimensional nanofriction in three-dimensional ion Coulomb crystals}

\author{L. A. Rüffert}
\affiliation{Physikalisch-Technische Bundesanstalt, Bundesallee 100, Braunschweig, 38116, Germany}

\author{T. E. Mehlstäubler}
\affiliation{Physikalisch-Technische Bundesanstalt, Bundesallee 100, Braunschweig, 38116, Germany}
\affiliation{Institut für Quantenoptik, Leibniz Universität Hannover, Welfengarten 1, Hannover, 30167, Germany}
\affiliation{Laboratorium für Nano- und Quantenengineering, Leibniz Universität Hannover, Schneiderberg 39, Hannover, 30167, Germany}

\renewcommand{\thesubsubsection}{\Alph{subsection}\arabic{subsubsection}}

\date{\today}

\begin{abstract}
Self-organized three-dimensional (3D) ion Coulomb crystals in linear Paul traps naturally form concentric shells that provide a curved, atomically resolved interface for studying two-dimensional (2D) nanofriction. Building on prior work that used 2D ion crystals to investigate one-dimensional (1D) nanofriction and orientational melting, we leverage this foundation to extend friction studies from linear ion chains and planar rings to 3D shell structures. 
Using molecular-dynamics simulations, we first map shell formation as a function of ion number $N$ and the trapping aspect ratio, yielding a simple relation that can aid ion-number estimation in experiments. We compute a Peierls–Nabarro-type potential for the rotation of the outer shell against the static inner core, by treating the angle of rotation as a collective coordinate. We find that changing $N$ by just one can alter the effective rotational barrier by up to a factor of $\sim7$, while in extreme cases changing $N$ by a few can modify the barrier by up to a factor of $\sim60$. We quantify a geometric commensurability measure and show that the barrier reflects a system-dependent interplay between the inter-shell interaction, the structural response of the outer shell and the confining potential of the trap.
Using dynamical simulations, we apply rotational torques to the outer shell and identify pinned, stick-slip, and smooth-sliding regimes whose depinning thresholds depend sensitively on ion number, inner-shell geometry, and trap aspect ratio, with some configurations exhibiting hysteresis due to torque-induced metastable states.
We find that spatially varying coupling to the inner-core corrugation can create coexisting fast and slow moving domains within the rotating outer shell, realizing multidimensional friction where intra-shell shear and inter-shell nanofriction act simultaneously. Our results highlight the utility of self-organized ion Coulomb crystals as model systems for 2D nanofriction and open up new possibilities for stabilizing complex systems and developing ultra-low-friction nanomechanical systems, such as ion-based nanorotors and torque sensors.
\end{abstract}

\maketitle

\section{Introduction}
\label{sec:Introduction}
Dry friction is a fundamental phenomenon that governs interactions across scales, from nanotribology to macroscale systems. Frictional losses due to tribological contacts account for nearly a quarter of global energy consumption and impact key industrial sectors such as transportation, power generation, and manufacturing~\cite{Holmberg_Friction_2017}. Understanding and controlling friction at the nanoscale is therefore of great importance. Various methods have been developed to investigate nanoscale friction, including atomic force microscopy (AFM) \cite{binnig_AtomicForceMicroscope_1986, martin_AtomicForceMicroscope_1987, Mate_AtomicscaleFrictionTungsten_1987}, tribological studies of two-dimensional (2D) materials like graphene \cite{dienwiebel_SuperlubricityGraphite_2004, xu_VanishingStickSlip_2011, Andersson_Friction_2020}, and model-friction experiments with trapped ions and colloids in optical lattices \cite{vanossiStaticDynamicFriction2012, mandelli_Nanofriction_2013, bylinskii_friction_2015, Kiethe_Nanofriction_2018, brazda_Aubry_2018}.

Trapped Coulomb crystals in Penning and Paul traps offer a powerful platform for studying nanofriction in highly controllable environments. These systems are used across various fields, including quantum computing \cite{cirac_QuantumComputationsCold_1995, blatt_EntangledStatesTrapped_2008, monroe_ScalingIonTrap_2013}, precision spectroscopy \cite{arnold_ProspectsAtomicClocks_2015, keller_EvaluationTrapinducedSystematic_2016, keller_MultiionSpectros_2024, schmidtSpectroscopyUsingQuantum2005, champenoisIonRingLinear2010}, and fundamental studies of quantum many-body phenomena \cite{islam_OnsetQuantumPhase_2011, britton_EngineeredTwodimensionalIsing_2012, bohnet_QuantumSpinDynamics_2016, Ruffert_Domain_2024}. Beyond these applications, the dynamics of large three-dimensional ion Coulomb crystals and strongly coupled non-neutral plasmas have been investigated extensively, establishing these systems as model platforms for collective dynamics and transport in the strongly coupled regime~\cite{dubin_Plasmas_1999,champenoisIonRingLinear2010,poindron2021thermal,baldovin_StronglyCoupled_2024,Zaris_simulations_2025}.

The static properties of ion Coulomb crystals have been widely studied. As particle number increases, the system forms concentric ring- or shell-like structures in 2D and 3D geometries, respectively \cite{gilbert_ShellStructurePhaseMagnetically_1988, Hasse_cylindrical_Coulomb_1990, hasse_StructureMadelungEnergy_1991, drewsen_large_1998, radzvilavicius_TopologicalDefectMotifs_2011, Drewsen_ICCs_2015}.
Recent work has investigated orientational melting in 2D crystals with up to $15$ ions, interpreted as thermally activated rotation of the outer ion ring around the central core \cite{duca_orientational_2023}. 
More generally, enhanced stability at specific particle numbers (“magic numbers”) has long been discussed in finite charged clusters, both in three-dimensional Coulomb crystals and in two-dimensional mesoscopic systems, where shell closures and highly symmetric configurations were linked to suppressed intershell rotation and melting~\cite{tsuruta_BindingEnergyMicrostructure_1993, schweigert_SpectralPropertiesClassical_1995, arp_3DCoulombBalls_2005, tomecka_MultistepRadialMelting_2005}.
On a similar note, studies on the Wigner crystallization of 2D electron clusters showed a strong dependence of the solid-to-liquid phase transition on the particle number~\cite{bonitz_SingleelectronControlWigner_2002, golubnychiy_ControllingIntershellRotations_2003}.

In this work, we employ molecular-dynamics simulations to investigate shell formation and inter-shell nanofriction in finite three-dimensional ion Coulomb crystals, using constrained and driven outer-shell rotation to show how ion number, trap aspect ratio, shell commensurability, and collective relaxation determine rotational barriers, depinning thresholds, and multidimensional sliding dynamics.

First, we find that the scaling for the number of shells in finite, spheroidal crystals is captured by a simple power-law, only depending on the ratio of the trapping potential $\alpha$ and particle number $N$, allowing simple estimations of ion numbers in experimental settings (Conclusion A). 

We then treat rotating shells in the view of nanofriction and calculate a Peierls-Nabarro-type potential of the outer shell rotation over the static inner core, by treating the angle of rotation as a collective coordinate.
We find that the effective barrier for shell rotation is highly sensitive to $N$, with changes by a factor of up to $\sim 7$, when $N$ is changed by one and by a factor of up to $60$ when $N$ is changed by a few. We quantify a geometric commensurability measure and find that while inter-shell commensurability provides an important contribution to these fluctuations, the full barrier is generally governed by a system-dependent interplay of inter-shell, outer-shell, and trap-related energy changes (Conclusion B).

When applying rotational torques to investigate dynamical friction between shells, we find that the depinning thresholds of the outer shell qualitatively follow the energy barriers obtained in Conclusion B. For some configurations, the driven system exhibits a hysteretic response due to torque-induced metastable states. Moreover, certain systems reveal a non-uniform distribution of the angular velocity along the axis of rotation, leading to 1D friction between ion segments within the rotating 2D shell and giving rise to complex, multidimensional friction phenomena (Conclusion C).

Our findings help to identify the mechanisms that result in a higher resilience to rotation of the outer shell which enhances the crystal stability against orientational melting, finding possible applications in multi-ion clocks \cite{keller_EvaluationTrapinducedSystematic_2016, arnold_ProspectsAtomicClocks_2015}, quantum simulators \cite{britton_EngineeredTwodimensionalIsing_2012, kiesenhofer_TwoDim_2023} and ion spectroscopy experiments \cite{schmidtSpectroscopyUsingQuantum2005, keller_MultiionSpectros_2024}. Conversely, structures with a low potential barrier for shell rotation might be of interest in the design of ultra-low-friction nanomechanical systems, such as ion-based nanorotors, gyroscopes or ultrasensitive torque detectors \cite{ahn_UltrasensitiveTorqueDetection_2020, ohira_2020, shao_MolecularRotorsDesigned_2020, singhania_AccountsAppliedMolecular_2023}.

The paper is structured as follows: In Sec.~\ref{sec:Theory} we first give an overview about the trapping of laser-cooled ions in a harmonic ion trap and the resulting self-organized Coulomb crystals, their shell structures in 3D configurations as well as an introduction over the different models of nanofriction. 

Following this, we outline our simulation methodology in Sec.~\ref{sec:methods} and discuss our results in Sec.~\ref{sec:results}, first by analyzing shell formation and its dependence on the particle number and trapping potential.

In Sec.~\ref{subsec:energy_barrier}, we calculate a Peierls-Nabarro-type potential of the outer shell rotation over the corrugation potential of the static inner shell by treating the rotation angle as a collective coordinate and analyze the effective barrier for outer shell rotation.

Using our findings, we apply a range of external torques to the outer shell of selected configurations and use the resulting angular velocity to identify different dynamical regimes, hysteresis and the formation of domains with varying angular velocity. These results are presented in Sec.~\ref{subsec:intershell_friction}.

Finally, we discuss possible experimental realizations of our findings in Sec.~\ref{subsec:exp_feasibility} and summarize the main results of this study in Sec.~\ref{sec:summary}.

\section{Theoretical background}
\label{sec:Theory}

\subsection{Ion Coulomb crystals}
Ions which are being trapped in a Paul or Penning trap and are laser-cooled to a few $\si{\milli\kelvin}$ form self-organized Coulomb crystals~\cite{dubin_Plasmas_1999}.
In a Paul trap, a rapidly oscillating rf electric field generates an average confining force on the ions that, in the ponderomotive approximation, is described by a time-independent quadratic potential in all three spatial directions~\cite{Paul1990}.

We consider $N$ identical ions with positions $\vec r_i = (x_i, y_i, z_i)$, mass $m$, and charge $Q$, interacting via the Coulomb force. Approximating the rf-potential to be harmonic, the total potential energy of the system can be written as
\begin{align} \label{eq:potential}
\mathcal{V} = \sum_i^N \frac{m}{2} \left( \omega_{x}^2 x_i^2 + \omega_{y}^2 y_i^2 + \omega_{z}^2 z_i^2 \right) + \sum_{i<j}^N \frac{Q^2}{4\pi \epsilon_0 d_{ij}},
\end{align}
where $\epsilon_0$ is the vacuum permittivity and $d_{ij} = \norm{\vec{r}_i - \vec{r}_j}$ the distance between ions $i$ and $j$.
The secular frequencies $\omega_x$, $\omega_y$, and $\omega_z$ define the confinement strength in each direction and thus the overall shape of the ion crystal. We define the axial direction (along which only static fields are applied) as the $z$-axis, and the radial direction in the $xy$-plane, so that the secular frequencies follow:
\begin{equation}
\begin{aligned}
\omega_{z}^2 &= \frac{Q}{m} u_\text{DC}, \\
\frac{\omega_{x/y}^2}{\omega_{z}^2} &= \frac{1}{2} \frac{Q}{m} \frac{u_\text{rf}^2}{u_\text{DC} \Omega_\text{rf}^2} - \frac{1}{2} \mp c_{xy}\,.
\end{aligned}
\label{eq:omegas}
\end{equation}
Here, $u_\text{DC}$ and $u_\text{rf}$ are the static and oscillating field gradients, $\Omega_\text{rf}$ is the rf drive frequency, and $c_{xy}$ accounts for anisotropies of the radial directions.

At zero temperature, the ions form a stable configuration determined by the balance between Coulomb repulsion and the confining potential. By adjusting the trap frequencies, the shape of the resulting crystal and the distances between the ions can be tuned. In this work, we focus on spheroidal crystals, defined by
\begin{equation}
\omega_z \le \omega_r,
\end{equation}
with degenerate radial frequencies $\omega_r = \omega_{x,y}$, which allows the overall geometry to be characterized by the aspect ratio
\begin{equation}
\alpha = \omega_r^2 / \omega_z^2.
\end{equation}
For $\alpha > 1$, the crystal elongates along the $z$-axis and compresses in the radial direction \cite{Okada_characterization_2010}.

\subsection{Shell formation in self-organized Coulomb crystals}
\label{sec:shell_formation_theory}
\begin{figure*}
    \includegraphics[width=0.8\textwidth]{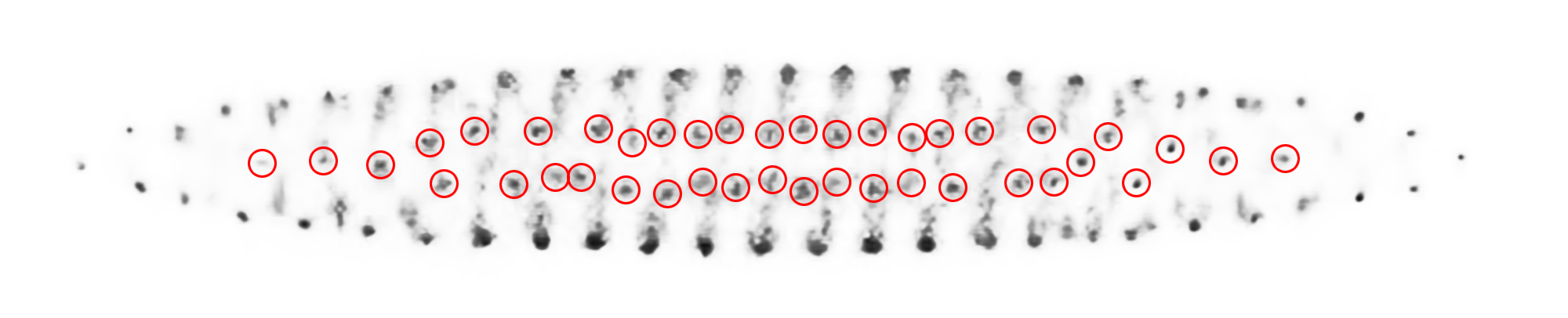}
    \caption{\justifying Experimental image of a three-dimensional (3D), Doppler-cooled Coulomb crystal of approximately 200 $^{172}$Yb$^{+}$ ions obtained in our laboratory using an EMCCD camera and shown here to illustrate the shell geometry and its experimental realization. The aspect ratio of the trapping potential is $\alpha\approx9$. The crystal consists of an inner shell (ions highlighted in red) and a helix shaped outer shell, which wraps around the inner structure.}
    \label{fig:experiment_crystal}
\end{figure*}
The emergence of shell structures in ion Coulomb crystals has been demonstrated both theoretically and experimentally on many occasions: Simulations with up to several thousand ions have been performed under symmetric trapping conditions ($\omega_x = \omega_y = \omega_z$)~\cite{Rahman_OCP_1986, Totsuji_plasma_physics_1987, Dubin_computer_simulation_IC_1988}, while the shell formation in infinitely long, cylindrically confined Coulomb systems has been studied analytically~\cite{Hasse_cylindrical_Coulomb_1990}. Monte Carlo simulations revealed that the dimensionless linear particle density
\begin{equation}
\lambda = \sigma a / Q,
\end{equation}
where $\sigma$ is the linear charge density and $a$ the Wigner-Seitz radius, determines the number of shells in such systems. Hasse and Schiffer found, that the number of shells approximately follows the empirical relation
\begin{equation}
\label{eq:N_shell_approx}
N_s \approx 0.78\,\sqrt\lambda
\end{equation}
for infinitely long Coulomb crystals~\cite{Hasse_cylindrical_Coulomb_1990}. Experiments in ring-shaped traps have confirmed this scaling behavior \cite{Birkl_multi_shell_ring_1992}. However, Coulomb crystals confined in linear Paul traps always exhibit finite boundaries along the axial direction of the trap.
Figure~\ref{fig:experiment_crystal} shows an experimental image from our laboratory of a 3D Coulomb crystal consisting of approximately 200 ${}^{172}$Yb$^+$ ions in a Paul trap, forming two distinct shells. The inner shell ions are highlighted in red, and the trapping potential has an aspect ratio of $\alpha \approx 9$.

In very large Coulomb crystals with prolate spheroidal shapes, the number of shells also approximately follows Eq.~\eqref{eq:N_shell_approx} \cite{drewsen_large_1998}. Our study on smaller systems ($N\leq 800)$ and varying values of $\alpha$ finds a good agreement with this scaling with a slightly adjusted prefactor of ${N_s = 0.83\,\sqrt{\lambda}}$. Although many studies have explored how the overall crystal shape depends on the ratio of the trapping potential \cite{Okada_characterization_2010, Drewsen_ICCs_2015}, less attention has been paid to how the number of shells scales with $\alpha$. The scaling of the number of shells with ion number and $\alpha$ will be discussed in the results section. 


\subsection{Nanofriction}
Two of the most widely used theoretical models to describe nanoscale friction are the Prandtl-Tomlinson (PT) and the Frenkel-Kontorova (FK) model. Both of which provide key insights into atomic-scale friction.

\medskip
\noindent\textbf{PT model: single particle dynamics} \\
The PT model \cite{Prandtl1928, Tomlinson1929} describes a single point mass, such as an Atomic Force Microscope (AFM) tip, moving over a periodic potential representing a crystal surface. A spring connects the tip to a carrier moving at constant speed. The energy of the system is made up of the periodic interaction potential and the elastic energy of the spring. This setup leads to stick-slip motion, where the tip remains trapped in local minima until the spring force exceeds a threshold, causing a sudden jump of the tip over the corrugation barrier, which is characteristic of nanoscale friction.

\medskip
\noindent\textbf{FK model: many-body dynamics} \\
The FK model \cite{FrenkelKontorova1938} extends the concept of a single particle on a periodic potential to a chain of harmonically coupled particles interacting with a periodic substrate. Originally developed to describe crystal dislocations, it also serves as a model for friction on the nanoscale.
A key prediction in systems with competing length scales is the Aubry transition, which separates an unpinned, superlubric phase from a pinned phase. For incommensurate ratios between the particle spacing and the substrate periodicity, the chain can slide without finite static-friction below a critical corrugation strength. Above this critical corrugation, the system becomes pinned and a finite force is required to initiate sliding \cite{aubry_ConceptTransitions_1978, peyrard_CriticalTransition_1983}.

Experimentally, such Aubry-type transitions have been demonstrated in well-controlled systems such as ion Coulomb crystals in rf Paul traps, 2D colloidal monolayers and also a self-organized crystal \cite{kiethe_probing_2017, Kiethe_Nanofriction_2018, brazda_Aubry_2018, vuletic_2020}.
While the Aubry transition is characterized by a sharp, well-defined change between a pinned and an unpinned sliding state, frictional motion in most macroscopic and inhomogeneous mesoscopic systems does not usually exhibit such a discrete transition. Instead, stick-slip motion often gradually evolves into smooth sliding as the driving velocity increases or the coupling between frictional layers is reduced \cite{voisin_LongTermFriction_2007, gourdon_TransitionsSmoothComplex_2003, drummond_DynamicPhaseTransitions_2001}.




\subsubsection{1D nanofriction}
\begin{figure}
    \includegraphics[width=\linewidth]{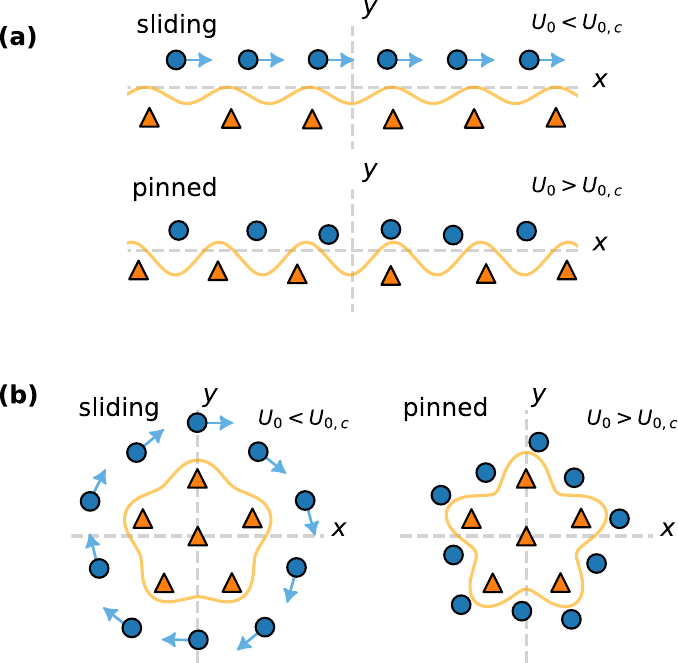}
    \caption{\justifying Illustrative example of one-dimensional (1D) nanofriction in neighboring ion chains. Ion positions are represented by blue circles and orange triangles. The substrate potential (orange line) created by the triangular-marked ions induces a corrugation, which resists the sliding motion of the blue ion chain. Arrows indicate the motion lateral to the corrugation layer. The corrugation depth $U_0$ determines the transition from a pinned state to a free sliding state. (a) Adjacent ion chains with different periodicities. A similar setup was experimentally realized using a topological defect in \cite{Kiethe_Nanofriction_2018} to verify an Aubry-type transition from sliding to pinning when the corrugation depth $U_0$ exceeds a critical value $U_{0,c}$. (b) Ring configuration of two adjacent ion chains. The number of ions directly dictates the periodicity between the two chains and therefore the commensurability. A similar system was used by Duca et al. to investigate orientational melting in 2D Coulomb crystals \cite{duca_orientational_2023}. We simulate the pinned state by  fixing the positions of the corrugation ions and increase the trapping potential, until the corrugation depth exceeds a critical value $U_{0,c}$, causing the system to be pinned.}
    \label{fig:1D_nanofriction}
\end{figure}
\noindent\textbf{External corrugation potential}\\
The Frenkel-Kontorova (FK) model can be emulated by moving a linear ion chain across an optical lattice, which acts as a static corrugation potential \cite{benassi_nanofriction_2011, braiman_Symmetrybreaking_1990, mandelli_Nanofriction_2013, bylinskii_friction_2015}. In this configuration, the ions are displaced via static electric fields while the lattice remains fixed. However, this setup only approximates realistic nanocontacts, where mutual interactions between atomic layers lead to deformations and backaction of the corrugation potential. 

\medskip
\noindent\textbf{Interacting ion chains} \\
Kiethe et al.~\cite{kiethe_probing_2017, Kiethe_Nanofriction_2018} demonstrated that Aubry-type transitions, soft modes and Hull functions can also be identified in self-organized systems of Coulomb crystals where the friction is caused by interacting ion chains. The work experimentally verified a sliding-to-pinning Aubry-type transition in a system with two adjacent ion chains, where a topological defect induced an incommensurability between the chains. The system is sketched in Fig.~\ref{fig:1D_nanofriction}(a). The defect and the finite size of the system lead to a symmetry breaking at a critical corrugation strength, controlled via the radial trapping potential. The transition marks the boundary between a sliding and a pinned state in self-organized systems.

Earlier studies of finite charged clusters had already connected intershell rotation and melting behavior to magic-number configurations and enhanced orientational stability~\cite{schweigert_SpectralPropertiesClassical_1995, schweigert_RadialFluctuationInducedStabilization_2000, tomecka_MultistepRadialMelting_2005}.
Duca et al.~\cite{duca_orientational_2023} investigated orientational melting between adjacent ion rings and found that the energy barrier for relative rotation diminishes for certain ion numbers.
To demonstrate that these results are fully connected to the concept of 1D nanofriction, a proof-of-concept simulation can be performed by fixing the inner core of ions and changing the radial confinement. With this, an Aubry-type transition at a critical confinement strength can be found, analogous to the findings of Kiethe et al. \cite{Kiethe_Nanofriction_2018, kiethe_probing_2017}, depending on the commensurability as illustrated in Fig.~\ref{fig:1D_nanofriction}(b).

\subsubsection{2D Nanofriction}
\begin{figure}
    \includegraphics[width=\linewidth]{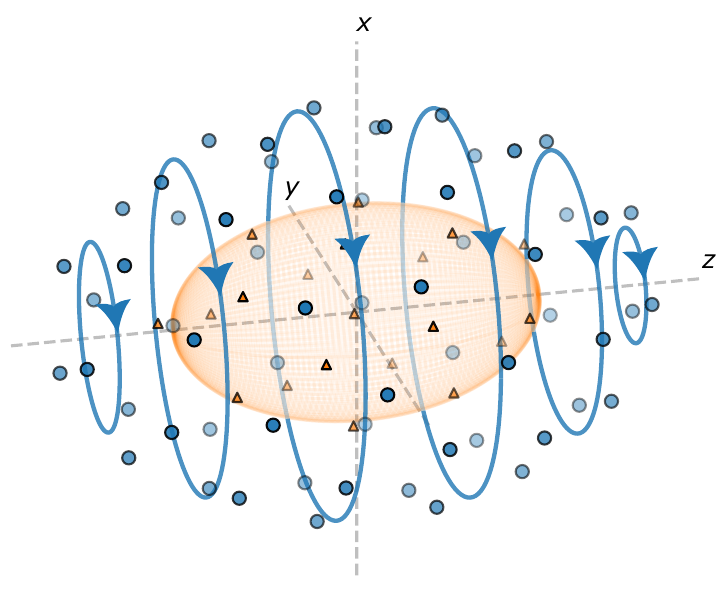}
    \caption{\justifying Model system of two-dimensional (2D) nanofriction in a three-dimensional (3D) self-organized Coulomb crystal ($N=80$) with two shells. Ions of the outer shell are represented by blue circles. Ions of the inner shell are shown as orange triangles. The ratio of the trapping frequencies is determined by $\omega_r >\omega_z$, causing the crystal to be of a spheroidal shape. The inner shell is approximated by a spheroidal surface (orange shade) for easier distinction of the two shells. The shear rotation around the $z$-axis of the outer shell is illustrated with blue circular arrows.}
    \label{fig:2D_nanofriction}
\end{figure}
One way to extend the concept of 1D nanofriction in self-organized systems to 2D is by analyzing the frictional coupling between concentric crystal shells in 3D Coulomb crystals, as shown in Fig.~\ref{fig:2D_nanofriction}. The inner shell (orange triangles), approximated by an ellipsoidal surface, acts as a 2D corrugation potential for the outer shell (blue circles). When the outer shell rotates (blue arrows), lateral forces arise due to this corrugation, giving rise to 2D nanofriction.

In an ideal 2D system, ions form a hexagonal lattice which is the structural ground state of an infinite 2D ion crystals \cite{Dubin_phase_transition_1993, Drewsen_ICCs_2015}. When two such lattices interact, the Frenkel-Kontorova (FK) model can be extended to describe interlayer friction in two dimensions \cite{Han_Superlubricity_2020, long_2D_FK_2010}. In smaller crystals, however, curvature of the layers and lattice defects complicate theoretical modeling. 
We adopt a phenomenological approach and use molecular dynamics (MD) simulations to identify configurations that exhibit notably different frictional behavior despite having similar ion numbers. 

Our analysis of the friction processes between shells is closely related to thermal angular melting scenarios discussed for finite Coulomb clusters. In both cases, the relevant low-energy degree of freedom is the relative rotational motion between neighboring shells. In this work, we probe this coordinate through a prescribed, driven rotation about a chosen axis and quantify its underlying corrugation. In contrast, the same relative motion emerges in orientational or intershell melting from unbiased thermal fluctuations and the associated loss of angular order. This connects the results of our work to the findings of Duca et al.~\cite{duca_orientational_2023} for adjacent ion rings and with the intershell-melting scenario identified by Apolinario and Peeters for isotropically confined 3D Coulomb clusters~\cite{apolinario_MeltingTransitions_2007}.


\section{Methods}
\label{sec:methods}
\subsection{Molecular dynamics simulations and parameters}
We describe the ion dynamics using the Langevin equation, which accounts for Brownian motion through a friction term with coefficient~$\eta$ and stochastic forces~$\vec{\epsilon}_i(t)$:
\begin{equation}
m_i\frac{d^2\vec r_i}{dt^2} = -\frac{d}{d\vec r_i}\mathcal{V} - m_i\eta\frac{d\vec r_i}{dt} + \vec \epsilon_i(t).
\end{equation}
The stochastic forces ensure thermal equilibrium at temperature~$T$, with their correlation structure determined by the fluctuation-dissipation theorem. In experiments, these forces originate from photon absorption and spontaneous emission in laser cooling. For Doppler cooling, the maximum friction coefficient is $\eta \propto \hbar k^2$ and shows best agreement with experimental data for ${\eta=(2.5 ... 3.0)\times 10^{-21}\si{\kilogram\per\second}}$ \cite{pyka_TopologicalDefectFormation_2013a}.
For finite temperatures, we simulate the system for several times the timescale $\eta^{-1}$ to ensure thermalization. The kinetic energy serves as a consistency check for the target temperature.
\label{subsec:shell_analsysis}
\begin{figure*}
    \includegraphics[width=\textwidth]{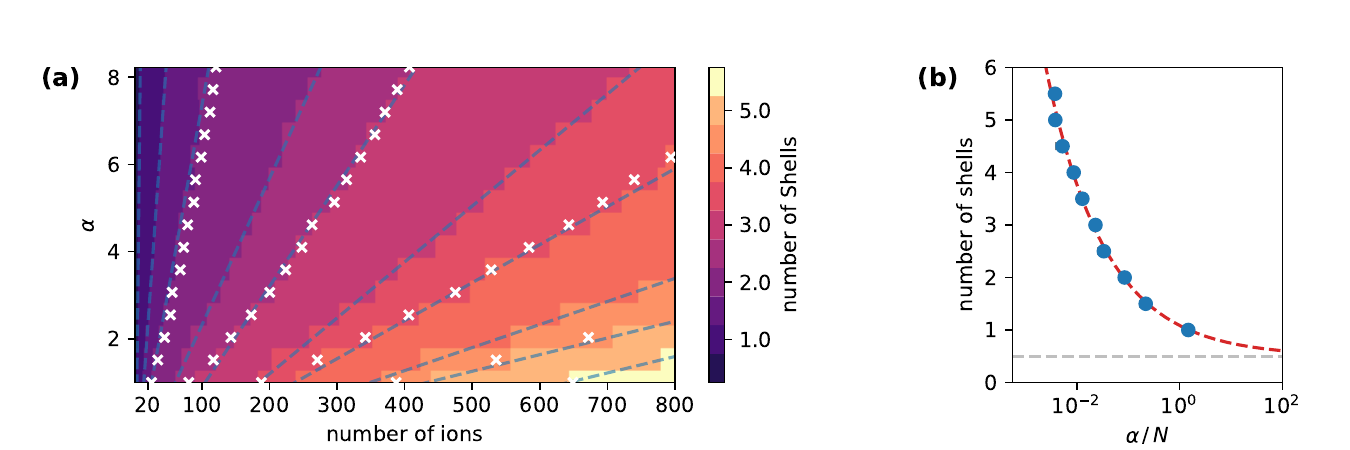}
    \caption{\justifying(a) Number of shells illustrated as different colored regions with respect to the number of ions $N$ and the aspect ratio of the trapping potential $\alpha=\omega_r^2/\omega_z^2$. A brighter color means a larger shell number. Half numbered shell counts describe crystals containing a linear chain of one or more ions on the $z$-axis. Linear fits to the transition regions are plotted as blue dashed lines. The white crosses mark the transition between full shell counts predicted by the linear particle density using $N_s = 0.83 \, \sqrt{\lambda}\,$. (b) The number of shells is plotted with respect to the ratio $\alpha/N$. A power-law function was fitted to the data (red dashed line). The gray dashed line shows the limit of $N_s=0.5$, which by definition represents the minimum achievable number of shells.}
    \label{fig:heatmap}
\end{figure*}

After thermalization, we record ion positions and velocities for further analysis. The choice of time step must account for the oscillation frequencies of the chosen ion species. As a continuation of our previous work \cite{Ruffert_Domain_2024}, we choose Be$^+$ with a mass of $9.01 \,\mathrm{amu}$ as the primary ion species in our simulations, although the results presented in this study are independent of the chosen ion species. We choose a fixed axial secular frequency of $\omega_z/2\pi \approx \SI{180}{\kilo\hertz}$ in all of the simulations. We find that setting the integration time step about a factor of 100 smaller than the fastest secular frequency in the simulated crystal sufficiently prevents numerical errors. The validity of the simulations have been verified using experimental results in previous work \cite{pyka_TopologicalDefectFormation_2013a, kiethe_probing_2017, Ruffert_Domain_2024}.
All molecular dynamics simulations presented in this work were performed using a custom code developed for laser-cooled ion Coulomb crystals, based on the impulse integrator for Langevin dynamics by Skeel and Izaguirre~\cite{skeel_ImpulseIntegratorLangevin_2002}. The code solves the Langevin equations of motion in the harmonic pseudopotential approximation and includes the full Coulomb interaction, Doppler-cooling damping and stochastic forces $\varepsilon(t)$, which are connected to the systems temperature by the fluctuation-dissipation relation
\begin{equation}
\langle \varepsilon_{\alpha j}(t)\, \varepsilon_{\beta i}(t')\rangle = 2\eta k_B T \delta_{\alpha \beta} \delta_{i j}(t-t')\, ,
\end{equation}
with $\alpha, \beta = x,y,z$~\cite{pyka_TopologicalDefectFormation_2013a, kiethe_probing_2017}.

It should be noted that the frictional behavior generally depends on the interplay between Coulomb interactions and external confinement. Within our static-harmonic pseudopotential model, a uniform rescaling of all secular frequencies leaves the dimensionless dynamics unchanged. We therefore fix $\omega_z \approx \SI{180}{\kilo\hertz}$ and vary the aspect ratio $\alpha = \omega_r^2/\omega_z^2$ and the ion number $N$, allowing us to explore structural rearrangements and their influence on friction.

\medskip
\noindent\textbf{Simulated annealing} \\
To consistently prepare our crystals in reproducible and comparable structural states, we employ a Simulated Annealing (SA) algorithm \cite{kirkSimulatedAnnealing1983, caracciolo_SA_2023}. Starting from randomized ion positions at a temperature where the crystal is molten, the system is cooled stepwise to $T=0$ while monitoring the potential energy. At each temperature step, the lowest-energy configuration sampled at that temperature is used as the candidate state for the next step. This complete cooling schedule defines a single SA run. Multiple SA runs are then performed from randomized and systematically altered initial conditions until the currently lowest-energy configuration has been recovered 10 times after its first occurrence. The complete algorithm is explained in more detail in Appendix~\ref{sec:ground state_prep}. 

While SA is a powerful algorithm to explore the energy landscape and reliably converges on low-energy configurations, it cannot guarantee that the true global minimum has been found. 
Throughout this work, the term ground state therefore refers to the lowest-energy configuration found reproducibly in repeated SA runs from randomized and systematically altered initial conditions. We cannot exclude the existence of other metastable configurations with slightly different energies or frictional properties, but the reproducibility of the selected configurations across repeated SA runs provides a robust practical basis for the analysis presented here.

\medskip
\noindent\textbf{Definition of the axis of rotation} \\
For the rotational analysis in Sec.~\ref{subsec:energy_barrier} and Sec.~\ref{subsec:intershell_friction}, we define the outer shell rotation about the $z$-axis. For ${\alpha>1}$, the crystal takes a spheroidal shape elongated along $z$ , and rotation about $z$ is the natural azimuthal shear mode between outer and inner shells. For the special case of $\alpha=1$, where no unique axis is selected by the trapping potential, the rotation axis is defined by the longest center-passing ion–ion line in the inner-shell. If no inner shell exists, the direction to the outer-shell ion farthest from the center is used instead. 

\section{Results}
\label{sec:results}
First, we will present the findings of the structural properties of finite Coulomb crystals, specifically the scaling of the number of shells with the particle number $N$ and the aspect ratio of the trapping potential $\alpha$. We will then focus on the effective energy barrier for the outer-shell rotation which will give us indications on the dynamic friction regimes which we will investigate in the last subsection of the results.

\subsection{Scaling of shell formation with ion number and aspect ratio of the trapping potential}
\label{subsec:N_shells_vs_alpha}
We examine the general formation of shell structures as a function of the particle number $N$ and the aspect ratio of the trapping potential $\alpha$. We define the point at which a new shell forms when the ion closest to the minimum of the trapping potential ($x=y=z=0$) is displaced from the $z$-axis, meaning
\begin{equation}
    x_i \ne 0 \;\land\; y_i \ne 0.
\end{equation}
Therefore, a string of ions forming along $z$ with $x=y=0$ would not be counted as a new shell. However, to allow for an even more nuanced analysis, we will define systems for which the ion closest to the minimum of the trapping potential fulfills the condition
\begin{equation}
    x_i = 0 \;\land\; y_i = 0,
    \label{eq:linear_shell_def}
\end{equation}
as a linear-shell structure. This distinction will allow for more precise estimations of particle numbers, based on the found number of shells. The exact method we use to evaluate the number of shells from a given set of ion positions is detailed in the Appendix~\ref{sec:shell_analysis}. 

The analysis is conducted over a broad parameter space, ranging from 1 to 800 ions and from aspect ratios of the trapping potential ranging from $\alpha=1.0$ to $8.22$. We calculate the structural ground state configuration for systems up to $100$ ions. For systems $N > 100$ we add individual ions to the outer shell along the $z$-axis one at a time from alternating directions and let the system equilibrate to limit computation time. This will most likely result in metastable crystal configurations for $N > 100$ that are of higher energy than the ground state. While the number of shells could, in principle, vary between metastable states and the ground state, extended simulations showed that such discrepancies have no significant impact on the overall trends in shell formation. Therefore, the study of shell structures in relation to $N$ and $\alpha$ remains robust, even when metastable states are considered. 

The findings of the analysis are given in Fig.~\ref{fig:heatmap}(a), where we plot the number of shells in dependence on the number of ions and the ratio of the trapping potential $\alpha$ as a color gradient. A brighter color indicates a larger number of shells. We also include the linear-shells by counting them as half shells (we will later account for this somewhat arbitrary choice by choosing an appropriate error estimate). We find a linear dependence of the transition regions between different numbers of shells on the number of ions and the aspect ratio of the trapping potential $\alpha$. Linear fits to the transition regions are shown as blue dashed lines in Fig.~\ref{fig:heatmap}(a). The slopes of these fits, which are the ratios of $\alpha / N$, become increasingly shallow for larger shell counts.

To compare the results to the empiric predictions for infinite cylindrical Coulomb crystals \eqref{eq:N_shell_approx}, made by Hasse and Schiffer \cite{Hasse_cylindrical_Coulomb_1990}, we estimate the linear density $\lambda$ for each system: Assuming that the linear density for the finite systems is roughly constant around a range $|z|<\Delta z$, we can approximate $\lambda$ by counting the number of particles within this range. Although the choice of $\Delta z$ is somewhat arbitrary, values for $\Delta z=0.1 \, L_z$ to $0.4 \, L_z$ have been tested, with $L_z$ being the length of the crystal in the $z$-direction, without having a significant impact on the final results. We find that the shell transitions follow the empirical relation of \begin{equation}
    N_s \approx 0.83 \, \sqrt{\lambda}\;,
\end{equation}
highlighted as white crosses in Fig.~\ref{fig:heatmap}(a), which is in good agreement with the empirical findings by Hasse et al. $N_s \approx 0.78 \, \sqrt{\lambda}\;$. For systems with $\alpha\approx1$ the linear density is slightly overestimated due to the more pronounced curvature of the spheroid along $z$, resulting in a larger offset to this empirical relation. However, the estimation of the number of shells based on the linear density generally holds up for smaller spheroidal shaped crystals with finite boundaries. It is important to note that for a fixed aspect ratio $\alpha$, a uniform rescaling of all trap frequencies preserves the dimensionless linear charge density $\lambda$, which primarily determines the number of shells in the crystal. The observed shell configurations are therefore invariant under changes of the absolute trapping strength, as long as $\alpha$ is kept constant. 

Following the results of Fig.~\ref{fig:heatmap}(a), we find that the number of shells follows a power-law relation of 
\begin{equation}
\label{eq:N_shell_vs_alpha}
    N_s \approx 0.6 \, \cdot \, \left(\frac{\alpha}{N}\right)^{-0.37}+0.5,
\end{equation}
which is shown as a red, dashed line in Fig.~\ref{fig:heatmap}(b). The blue data points show the slopes $\alpha/N$ of the linear fits (blue dashed lines in Fig.~\ref{fig:heatmap}(a)). To account for the fact, that the definition of the half-shell structure remains somewhat arbitrary, we apply an additional error estimate of $20\%$ to the ratio of $\alpha/N$ for non-integer shell numbers. The curve approaches the limit value of $0.5$, which by definition \eqref{eq:linear_shell_def} represents the minimum achievable number of shells.
\begin{figure*}
    \includegraphics[width=\textwidth]{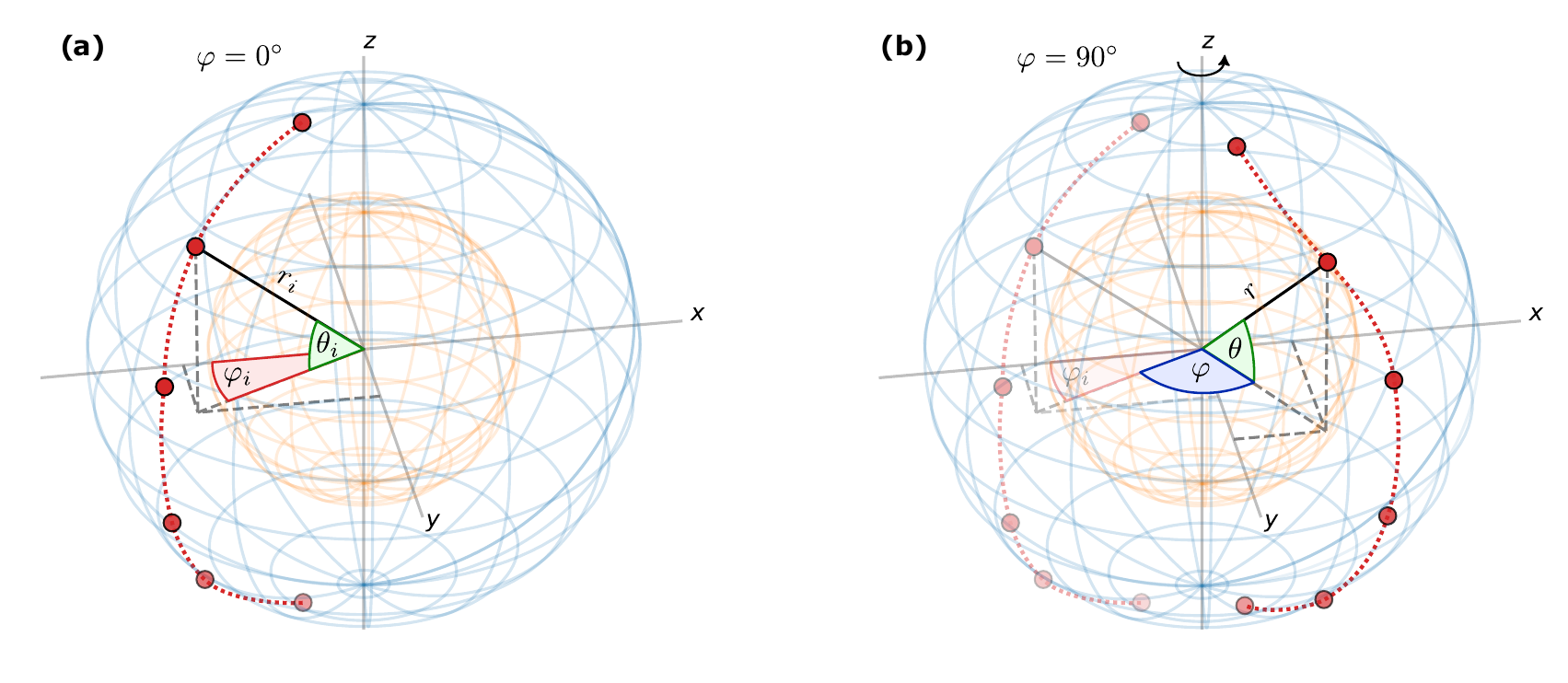}
    \caption{\justifying Rotation of the outer shell (blue wireframe) around the $z$-axis for an $N=80$ ion crystal with trapping aspect ratio $\alpha = 1$. Inner shells are shown in orange. For clarity, only the driving ions (red circles) are plotted explicitly and are connected by a dashed red line for easier visualization. The position of each driving ion is described in spheroidal coordinates $(r, \theta, \varphi)$, where $\varphi$ is the azimuthal angle relevant for the rotation. (a) Initial configuration at $\varphi = 0^\circ$. The initial coordinates $r_i$, $\theta_i$, and $\varphi_i$ of a sample ion are shown in black, green, and red, respectively. (b) Configuration after rotation to $\varphi = 90^\circ$. The initial position ($\varphi_i$) is shown with reduced opacity for reference. The rotation angle $\varphi$ is measured relative to $\varphi_i$ and is identical for all driving ions. The outer shell is rotated by incrementally rotating the driving ions (red markers) around the $z$-axis in $\Delta \varphi = 1^\circ$ steps. While radial and polar motion is allowed during relaxation, for the driving ions, the azimuthal position is fixed during each rotation step. The dynamics of all other ions of the outer shell are unrestricted. The ions of the inner shells are kept static throughout the simulation.}
    \label{fig:shell_rotation}
\end{figure*}

Since the number of shells can in general be easily identified by modern EMCCD imaging systems in Paul-trap experiments and the aspect ratio of the trapping potential is given by the applied dc- and rf-voltages, the number of ions in an experiment can be estimated by using relation \eqref{eq:N_shell_vs_alpha}.

\medskip
\noindent\textbf{Conclusion A} \\
We mapped the number of shells versus ion number $N$ and trap aspect ratio $\alpha$ and observe near-linear transition boundaries between different shell-regimes whose slopes $\alpha/N$ decrease with higher shell counts. We find that the transitions between shell numbers follows the local linear-density scaling $N_s \approx 0.83\sqrt{\lambda}$ (consistent with the results of Hasse and Schiffer for infinite cylindrical Coulomb crystals), and we find a power-law scaling of the number of shells with the ratio $\alpha/N$ (see Eq.~\eqref{eq:N_shell_vs_alpha}).
Since $N_s$ is easily imaged and $\alpha$ is known experimentally, these relations enable a general upper and lower bound estimate of $N$ in the experiment.

\subsection{Collective effects on inter-shell rotation}
\label{subsec:energy_barrier}
\begin{figure*}
    \includegraphics[width=0.8\textwidth]{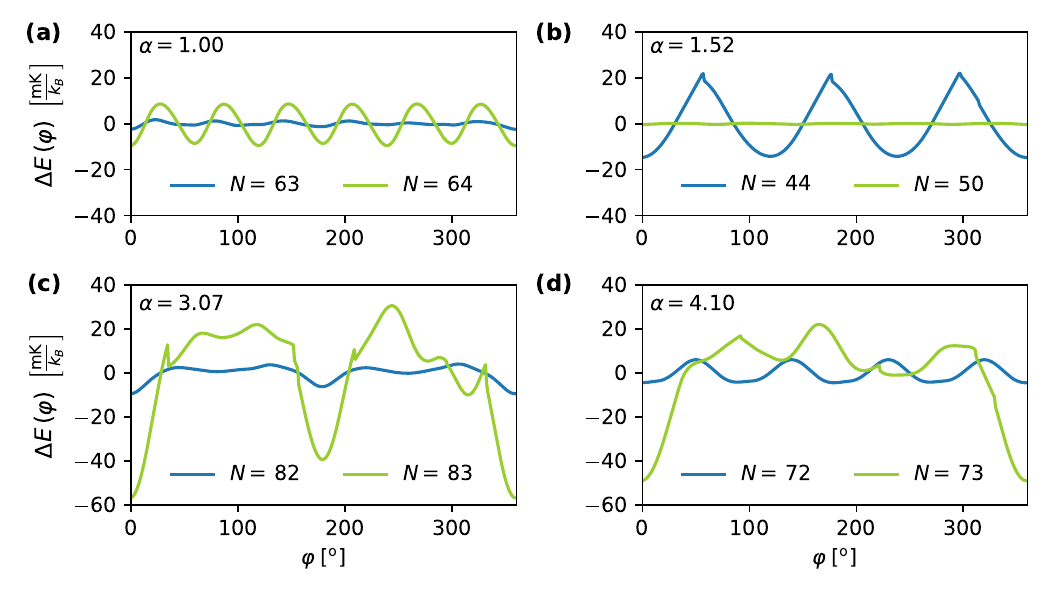}
    \caption{\justifying Angle dependent modulations of the Peierls-Nabarro-type potential of the outer shell for selected crystal configurations. The ratio of the trapping potential $\alpha=(1.0, 1.52,\, 3.07,\, 4.10)$ is indicated in each plot. The inner shell ions are kept static throughout the rotation. Periodicities in the different graphs reflect the structure of the inner shell. Non-differentiable points in $\Delta E(\varphi)$ indicate a temporary rearrangement of the outer shell ions during the rotation. The systems were selected based on the significant change in the energy barriers while only slightly changing the particle numbers between the respective systems.}
    \label{fig:E_pot_vs_theta}
\end{figure*}
\label{subsec:shell_rotation}
\begin{figure*}
    \includegraphics[width=0.9\textwidth]{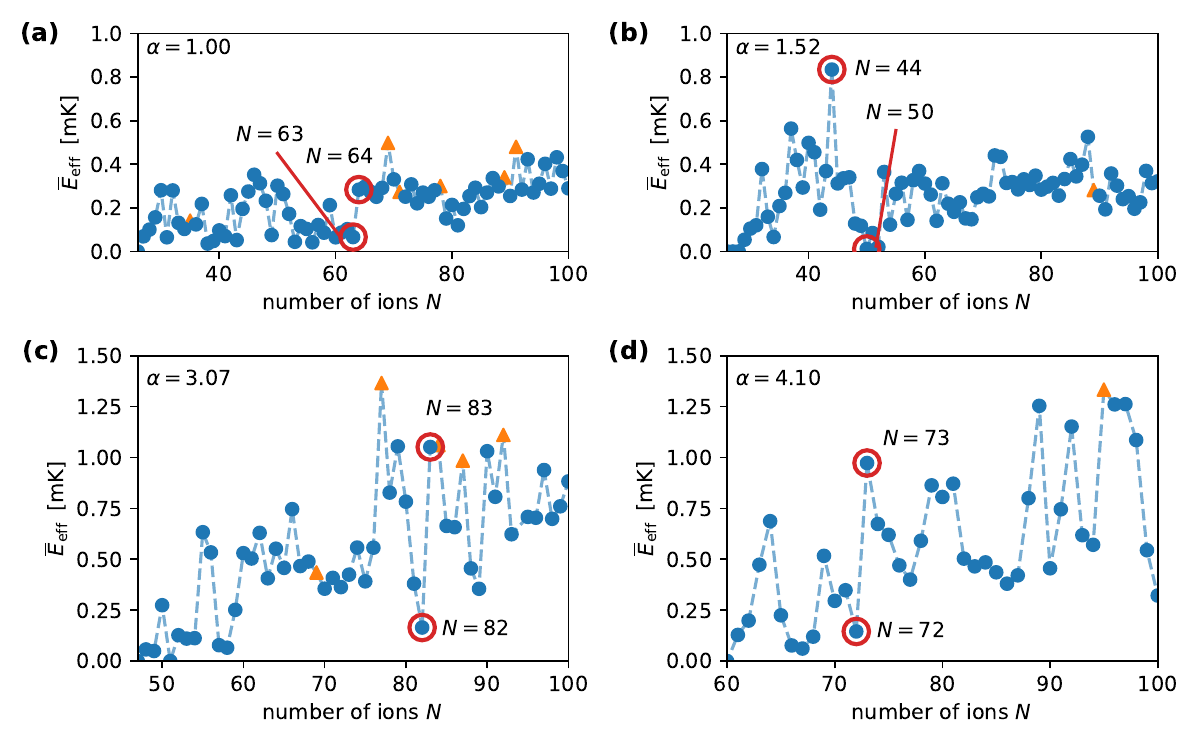}
    \caption{\justifying Normalized effective energy barrier $\overline{E}_\text{eff}$ for four selected ratios of the trapping potential ${\alpha=(1.0, 1.52,\, 3.07,\, 4.10)}$ plotted against the number of ions $N$. The ground state configuration was determined for each crystal. Selected configurations of interest are marked in red, where a large change in the effective energy is highlighted (corresponding to the Peierls-Nabarro-type potential in Fig.~\ref{fig:E_pot_vs_theta}). Orange triangles mark systems that did not return to their initial configuration after a full rotation of the outer shell due to irreversible reordering of the ions. Note that configurations where the inner shell only consists of a single particle or a string of ions has been neglected due to an effective barrier of $E_\text{eff}=0$.}
    \label{fig:eff_barrier}
\end{figure*}
The effective energy barrier provides direct insight into the frictional properties of sliding interfaces and can generally be defined as
\begin{equation}
E_\text{eff} = \max \Delta E(\vec{r}) = \max E(\vec{r}) - \min E(\vec{r}),
\end{equation}
where $E(\vec{r})$ is the total potential energy at a given position $\vec{r}$. We compute $E_\text{eff}$ for ground state configurations with up to 100 ions and selected trap aspect ratios $\alpha = (1, 1.52, 3.07, 4.10)$. Only crystals with at least two shells are considered ensuring inter-shell friction can occur. 

To map the potential energy landscape, we rotate the outer shell around the $z$-axis in discrete steps of ${\Delta \varphi = 1^\circ}$ and measure the potential energy $E(\varphi)$. During this process, the inner shell remains fixed, enforcing a shear motion between the shells. To drive the outer shell while allowing for minimization of the potential energy at each step, we first determine a continuous ion chain between the $z$-poles of the crystal that will drive the rotation. 

Fig.~\ref{fig:shell_rotation}(a) shows the initial setup for a $N=80$ ion crystal at $\alpha=1.0$. The driving ions which form the shortest connected chain between the poles along the $z$-axis are highlighted as red circles. For clarity, we only show wireframe representations of the outer shell (blue) and the inner shell (orange). The position of each driving ion is given in spheroidal coordinates $(r, \theta, \varphi)$, where $\varphi$ is the azimuthal angle relevant for the rotation.

The initial configuration shown in Fig.~\ref{fig:shell_rotation}(a) at $\varphi = 0^\circ$ defines the initial spheroidal coordinates $r_i$, $\theta_i$, and $\varphi_i$. Fig.~\ref{fig:shell_rotation}(b) shows the crystal after rotation of the outer shell to $\varphi = 90^\circ$. The rotation angle $\varphi$ is measured relative to $\varphi_i$ and the same angle $\varphi$ is applied to all driving ions, which push the rest of the outer shell by Coulomb repulsion. After each $\Delta \varphi$ step, the system relaxes to a local energy minimum, with the driving ions fixed at $\varphi$, while all other outer shell ions are free to move. After relaxation, the total potential energy of the system $E(\varphi+\Delta\varphi)$ is measured. 

The relaxed total energy $E(\varphi)$ plays the role of an effective Peierls–Nabarro-type (PN) potential of the rotational coordinate. Here the collective object is the entire outer shell treated as a rotated interface with fixed rotational angle $\varphi$ for the set of driving ions, while all remaining degrees of freedom are relaxed under this constraint. Because $E(\varphi)$ is evaluated after relaxation of the outer shell at each rotation step, it does not only reflect the direct interaction between the shells, but also contains contributions of the mutual interaction within the outer shell, and the trap-related energy cost along the constrained rotational path. A decomposition of these terms is discussed in more detail in Appendix~\ref{sec:dynamic_commensurability}. $E(\varphi)$ therefore captures the full energetic cost of relative shell motion in a finite mesoscopic system, where tangential sliding is coupled to structural relaxation of the outer shell. 
The trapping potential provides a restoring potential that stabilizes the crystal and constrains the relative motion of the shells. More generally, frictional motion in finite systems not only depend on the lateral corrugation between the sliding parts, but also on the stabilizing contributions that maintain their mechanical integrity and mutual confinement, even if their microscopic origin differs from the harmonic confinement used here.

After a full rotation of $360^\circ$, we calculate the angle dependent energy difference
\begin{equation}
    \Delta E(\varphi) = E(\varphi)-\langle E \rangle\,,
\end{equation}
where $\langle E \rangle$ is the average of the total potential energy $E(\varphi)$ taken over the full $360^\circ$ rotation. Examples of $\Delta E(\varphi)$ are shown for selected cases in Fig.~\ref{fig:E_pot_vs_theta}. We specifically chose systems that show significant differences in the potential energy landscape while being similar in ion number.

The selected systems in Fig.~\ref{fig:E_pot_vs_theta} fall into two main structural scenarios: for $\alpha=1.0 \;\; (N=63, 64)$ and $1.52 \;\; (N=44, 50)$, the inner shell remains nearly unchanged when $N$ is varied, whereas for $\alpha=3.07 \;\; (N=82, 83)$ and $4.10 \;\; (N=72, 73)$ the inner shell changes its helical arrangement. Accordingly, strong changes in $E(\varphi)$ can arise either from a reorganization of the outer shell at nearly fixed inner-shell geometry or from an actual restructuring of the inner shell.
The structure of the inner shells of the selected configurations is shown in the Appendix in Fig.~\ref{fig:inner_shell_configs}, which provides a reference for the following discussion.
The system $\alpha=1.0$, $N=64$ as well as $\alpha=1.52$, $N=44$ and $\alpha=4.10$, $N=72$ all exhibit periodic modulations in $E(\varphi)$ with periodicities of $60^\circ$, $120^\circ$ and $90^\circ$, respectively. These reflect the inner shell symmetries: In the $\alpha=1.0$ case the inner shell forms two hexagonal cells along the $z$-axis, causing a modulation of precisely $60^\circ$ in $E(\varphi)$. 

In the $\alpha=1.52$, $N=44$ case, the inner shell consists of two counter-rotated tetrahedra causing a $120^\circ$ modulation. In the $\alpha=4.10$, $N=72$ system, the inner shell forms a double helix of 10 ions with $90^\circ$ separation between ion pairs.

Systems with higher ion numbers at $\alpha=3.07$ and $\alpha=4.10$ show a series of non-differentiable points where $E(\varphi)$ abruptly changes its slope, indicating reordering of the outer shell during relaxation. Despite these reordering events, both systems return to their initial state after a full $360^\circ$ rotation, as $E(0^\circ) = E(360^\circ)$. This symmetry is not always preserved, as later results will show.



The selected systems demonstrate that small changes in ion number can significantly alter $E(\varphi)$, even when the shell structure changes only slightly. For example, for the two $\alpha=1.0$ systems as well as both $\alpha=1.52$ systems, the respective inner shells remain unchanged between the different ion numbers, while the outer shell is slightly rearranged, leading to a modified relaxed rotational path and a significantly flatter energy landscape.

In the $\alpha=3.07$ case, the $N=83$ system contains two additional ions in the double-helix inner shell compared to $N=82$, thereby changing the shell geometry and the interactions that shape the energy landscape. For $\alpha=4.10$, the helical symmetry observed in the $N=72$ crystal is broken by an added ion in the $N=73$ system, distorting the periodicity in $E(\varphi)$ and raising the energy barrier for rotation. A more detailed analysis of the inner shell structures is given in Appendix~\ref{sec:inner_shell_structure}. 

We define the effective energy barrier as the peak-to-peak variation of the quasi-static Peierls–Nabarro-type energy landscape for the constrained rotation of the outer shell relative to the inner core,
\begin{equation}
E_\text{eff} = \max{E(\varphi)} - \min{E(\varphi)}.
\end{equation}
We normalize $E_\text{eff}$ by the particle number $N$ 
\begin{equation}
\overline{E}_\text{eff} = E_\text{eff} / N
\end{equation}
to account for an increase in potential energy with growing crystal size. We plot the results of $\overline{E}_\text{eff}$ for crystal ground states up to $N=100$ ions for trapping aspect ratios $\alpha=(1.0, 1.52, 3.07, 4.10)$ in Fig.~\ref{fig:eff_barrier}. We only plot data for crystals with at least 2 shells. Orange triangles highlight configurations, where $E(0^\circ) \ne E(360^\circ)$ due to irreversible reordering in the outer shell.

We find distinct local minima and maxima in $\overline{E}_\text{eff}$ for each system. 
The selected configurations of Fig.~\ref{fig:E_pot_vs_theta} are highlighted with red circles. We find that the absolute value of the effective energy barrier changes by a factor of $4.35$ in the highlighted $\alpha=1.0$ systems, by a factor of $6.46$ for $\alpha=3.07$ and by a factor of $6.78$ in the $\alpha=4.10$ system while the ion number only changes by $1$. In the case of $\alpha=1.52$, the change in the effective energy barrier is even more dramatic: here, the ion number differs by $6$, resulting in a change of the effective energy barrier by a factor of $61.65$.
We also compared $E_\text{eff}(N)$ to the second energy difference $\Delta_2E(N)$, which is a common measure of finite-size stability and magic-number effects in ion clusters~\cite{ludwig_StructureSpherical_2005, arp_3DCoulombBalls_2005, apolinario_StructuralDynamicalAspects_2007, apolinario_MeltingTransitions_2007, apolinario_MultipleRings3D_2008}. A detailed analysis of this comparison is given in Appendix~\ref{sec:second_energy_comparison}. We find no correspondence between the local extrema of $\Delta_2E(N)$ and those of $E_\text{eff}(N)$, indicating that the strong particle-number dependence of the rotational barrier is not primarily controlled by global cluster stability. 

A detailed energy decomposition together with a geometric commensurability analysis along the same relaxed rotational path is presented in Appendix~\ref{sec:dynamic_commensurability}. It shows that variations in inter-shell commensurability can account for an important part of the barrier fluctuations in some systems, but that $E_\text{eff}$ more generally reflects a system-dependent balance between inter-shell, outer-shell, and trap-related energy changes.

We also want to highlight the findings of Apolinario and Peeters~\cite{apolinario_MeltingTransitions_2007} who identified inter- and intrashell melting transitions in isotropically confined Coulomb clusters. In their work, the $N=38$ two-shell cluster was identified as having a thermally activated intershell melting transition at much lower temperatures compared to the intrashell melting. We find a pronounced minimum for the isotropic system $\alpha=1.0$, $N=38$ which is consistent with their finding. We stress, however, that the melting analysis probes thermally activated relative shell motion without a prescribed rotational axis, whereas $E_\text{eff}$ is obtained from a driven rotational path about a chosen axis. The agreement should therefore be understood as qualitative evidence that the intershell rotational degree about a specified axis is mechanically soft in our barrier analysis and thermally soft in their melting picture.

\medskip
\noindent\textbf{Conclusion B} \\
Our findings reveal a strong dependence of the effective rotational barrier on the ion number and the trap aspect ratio $\alpha$. 
Periodic features in the Peierls-Nabarro-type potential of the outer shell $E(\varphi)$ (see Fig.~\ref{fig:E_pot_vs_theta}) are linked to highly symmetric inner shell structures, while abrupt changes or $E(0^\circ) \ne E(360^\circ)$ indicate rearrangements during rotation. Low effective energy barriers indicate the systems susceptibility to reduced friction between shells and therefore orientational melting. We identified systems where the effective energy barrier of the outer shell rotation changes by a factor of up to $\sim7$ when changing the ion number $N$ by one and up to $\sim 62$ when changing $N$ by a few. A quantitative analysis (Appendix~\ref{sec:dynamic_commensurability}) shows that the rotational barrier is a composite quantity that depends on the interplay between inter-shell coupling, the geometric commensurability between inner and outer shell, and the structural response of the outer shell, including its relaxation in the trap potential. In addition, certain configurations exhibit irreversible reordering of the outer shell during rotation, suggesting that these structures follow energetically favorable pathways that do not return to their initial configuration. Our findings demonstrate that small changes in crystal structure and ion number can drastically alter the inter-shell potential barrier which has direct implications on the frictional behavior between shells.

\subsection{Frictional regimes and domain formation in rotating inhomogeneous Coulomb crystals}
\label{subsec:intershell_friction}
\begin{figure}
    \includegraphics[width=0.5\textwidth]{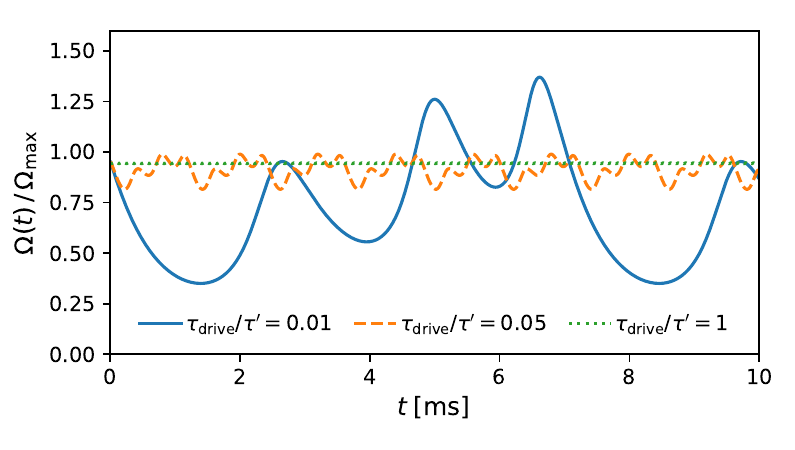}
    \caption{\justifying Angular velocity $\Omega(t)$ of the outer shell over time, normalized by the maximum angular velocity $\Omega_\text{max}$ achievable in the absence of corrugation-induced friction. The results shown are for the $N=44$ system with an aspect ratio of the trapping potential of $\alpha=1.52$. Results for three different driving torques are shown with the system being in a steady rotational state. Periodicities in the graphs indicate a full revolution of the outer shell. Oscillations indicate stick-slip motion caused by the corrugation potential. The average normalized angular velocity approaches $1$ for increasing driving torques, showing a continuous transition towards the smooth sliding regime.}
    \label{fig:ang_vel_vs_t}
\end{figure}
\begin{figure*}
    \includegraphics[width=0.8\textwidth]{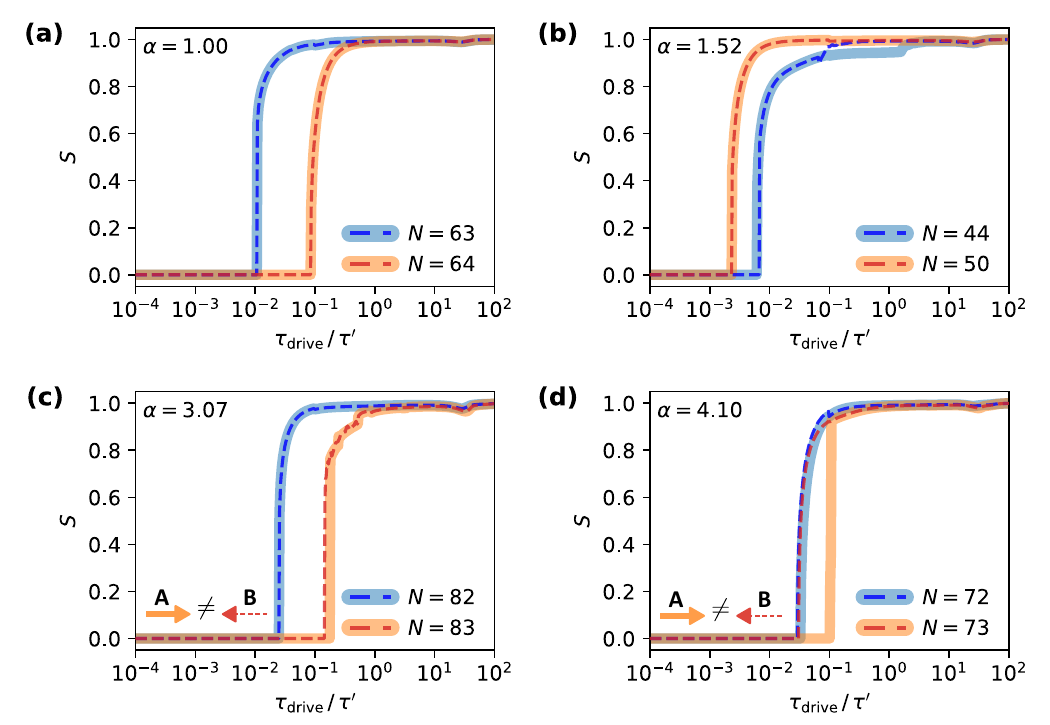}
    \caption{\justifying Sliding efficiency $S$ as a function of the driving torque for selected systems. The applied driving torque $\tau_\text{drive}/\tau'$ drives the rotation around the $z$-axis across the inner shells' corrugation potential. If none of the ions rotate by at least $1^\circ$ within $10$ $\si{\milli\second}$, the average velocity is set to zero and the simulation is terminated. Solid lines (bright blue and orange) show the response for increasing torque, dashed lines (dark blue and red) for decreasing torque to emphasize hysteresis. At a critical point, the transition from pinning to stick-slip happens. At high torques, $S$ asymptotically approaches $1$, indicating smooth sliding. The hysteresis for the $N=83$ system in (c) and the $N=73$ system in (d) stems from a reconfiguration of the outer shell structure when the system is initialized at high driving torques. During the adiabatic down-sweep, the system ends up in a metastable configuration \textbf{B} that differs from the groundstate configuration \textbf{A} (directionality of the torque-sweep is indicated by the colored arrows). See main text for details.}
    \label{fig:stick_slip_friction}
\end{figure*}
To investigate the inter-shell friction of the crystal configurations chosen in Sec.~\ref{subsec:shell_rotation}, we apply a constant rotational torque to the outer shell and  measure the resulting angular velocities, highlighting different frictional regimes and the dependence of the threshold-torque to initiate rotation on the ion number. 

The simulations employ Langevin dynamics, accounting for both the corrugation-induced lateral torque from the inner shell and stochastic forces with damping representing laser cooling. As a result, the constant driving torque $\tau_\text{drive}$ must overcome both the corrugation torque $\tau_\text{corr}$ and the damping torque $\tau_\text{damp}$ to initiate rotation. Once the rotation has reached a steady state, it must follow that
\begin{equation}
    \tau_\text{drive} + \langle\tau_\text{damp}\rangle + \langle\tau_\text{corr}\rangle = 0.
\end{equation}

In the following results, we will normalize all values for the torque by the factor
\begin{equation}
    \tau'=m_0 l_0^2 \omega_z^2,
\end{equation}
where $m_0$ is the mass of the chosen ion species and $l_0$ is the systems length scale defined as
\begin{equation}
    l_0 = \frac{e}{\sqrt{4\pi\epsilon_0m_0\omega_z^2}},
\end{equation}
to allow for an easier comparison between the different systems. 

To identify different frictional regimes, we will monitor the absolute value of the angular velocity $\Omega$ of the outer shell, given by
\begin{equation}
\Omega(t) = \frac{L(t)}{I(t)},
\end{equation}
where $L(t)$ denotes the time-dependent angular momentum and $I(t)$ the time-dependent moment of inertia of the outer shell.

Assuming steady-state balance between the applied torque and the damping torque due to laser cooling, the maximum angular velocity of the outer shell $\Omega_\text{max}$ is given by

\begin{equation}
    \Omega_\text{max}=\frac{\tau_\text{drive}}{\eta \; \langle I\rangle},
\end{equation}
where $\eta$ is the damping coefficient of the modeled laser cooling. Note, that while we use the time averaged value $\langle I \rangle$, slight fluctuations in position of the ions on the outer shell have negligible impact on the value of $\Omega_\text{max}$. 

Fig.~\ref{fig:ang_vel_vs_t} shows $\Omega(t) / \Omega_\text{max}$ over time, of the outer shell of the $\alpha=1.52$, $N=44$ configuration. Different driving torques are applied and the system is propagated until a steady state has been reached, before the data is recorded. The oscillations in $\Omega(t)/\Omega_\text{max}$ are indicative of stick-slip motion, while the periodicities in the graphs reflect full revolutions of the outer shell. For increasing driving torques, the impact of the corrugation potential on the sliding velocity gradually diminishes, and the system approaches the smooth sliding regime.

We now normalize the time averaged angular velocity by the maximum angular velocity $\Omega_\text{max}$ and define a sliding efficiency
\begin{equation}
S = \frac{\langle\Omega\rangle}{\Omega_\text{max}}\;,
\end{equation}
ranging from zero (pinned outer shell) to unity (smooth sliding) and will serve as the parameter to identify different sliding regimes.

We focus on the systems highlighted in Fig.~\ref{fig:eff_barrier} to analyze different frictional regimes and measure $S$ for a wide range of applied driving torques. We find that for driving torques exceeding $\tau_\text{drive}/\tau' \approx 100$, the centrifugal forces on the outer shell can cause structural reordering. We therefore only apply driving torques, for which the crystal structure remains stable, ranging from $\tau_\text{drive}/\tau' = 10^{-4}$ to $100$.
The results of our friction analysis are presented in Fig.~\ref{fig:stick_slip_friction}. 
\begin{figure}
    \includegraphics[width=0.48\textwidth]{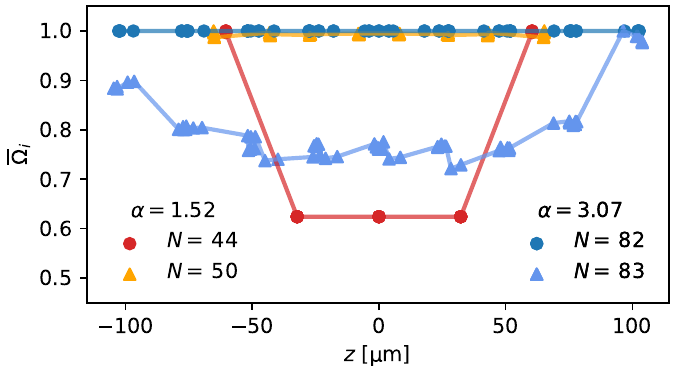}
    \caption{\justifying Normalized average angular velocity of each outer shell ion, $\overline{\Omega}_i$, plotted versus the ions' $z$ positions. For all data, the driving torque was chosen such that the outer shell sliding efficiency is $S=0.9$ (see Fig.~\ref{fig:stick_slip_friction}). Note, that ions lying close to the $z$-poles were excluded from this evaluation due to inconsistent values of $\overline{\Omega}_i$. We show the distributions only for $\alpha=1.52$ and $\alpha=3.07$, since the other cases ($\alpha=1.0$ and $\alpha=4.10$) yield nearly flat $\overline{\Omega}_i$ profiles for both ion numbers. For each pair of systems shown (differing only in ion number $N$), the $z$-dependence of $\overline{\Omega}_i$ is strongly $N$-dependent. 
    This contrast arises from system-dependent changes in the  rotational trajectories of the outer-shell ions when $N$ is varied (see main text for details).
    }
    \label{fig:ang_vel_vs_z}
\end{figure}
\begin{figure}
    \includegraphics[width=0.35\textwidth]{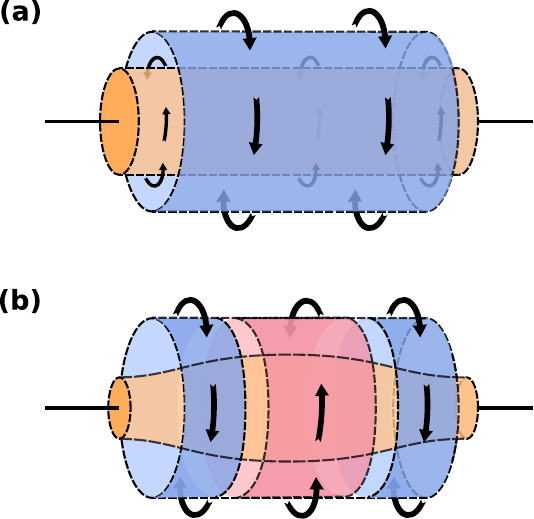}
    \caption{\justifying Schematic illustration of multidimensional friction in a self-organized system. Domains of freely moving particles are indicated by shaded areas, while different colors denote regions with distinct shear motion that generate friction at their interfaces.
    (a) The orange-shaded inner cylindrical core acts as a two-dimensional corrugation potential for the blue-shaded outer shell. Shear motion between these two surfaces gives rise to 2D inter-shell friction.
    (b) Spatial variations in the coupling strength to the corrugation potential of the inner core (indicated by the varying shape of the inner orange corrugation) induce spatially different friction strengths for the outer shell. The resulting shear between domains within the outer shell (blue and red shaded areas) gives rise to 1D intra-shell friction.}
    \label{fig:multi_dim_friction}
\end{figure}

All data has been recorded with the system being in a stationary rotation state. If no ion on the outer shell has rotated by at least $1^\circ$ within $\SI{10}{\milli\second}$, the simulation is canceled and $S$ is set to $0$. To investigate the systematic hysteresis, we also measured $S$ at each driving torque by initializing the system with a maximum driving torque of $\tau_\text{drive}/\tau'=100$ and adiabatically decrease the driving torque until a steady state has been reached at the target torque. The data for the hysteresis is plotted as dashed lines (blue and red) in Fig.~\ref{fig:stick_slip_friction}.

\begin{table}[h!]
\centering
\begin{tabular}{c @{\hspace{1em}} c @{\hspace{1em}} c @{\hspace{1em}} c}
\hline
$\alpha$ & $N$ & $\tau_\text{depin}/\tau'$ & $\tau_\text{pin}/\tau'$ \\
\hline
1.00 & 63 & 0.011 & 0.011 \\
1.00 & 64 & 0.085 & 0.085 \\
1.52 & 44 & 0.006 & 0.007 \\
1.52 & 50 & 0.002 & 0.002 \\
3.07 & 82 & 0.025 & 0.025 \\
3.07 & 83 & 0.181 & 0.142 \\
4.10 & 72 & 0.033 & 0.030 \\
4.10 & 73 & 0.104 & 0.031 \\
\hline
\end{tabular}
\caption{\justifying Depinning and pinning torques for different values of $\alpha$ and particle numbers $N$. Standard errors of the given torques are below 2\%.}
\label{tab:depinning}
\end{table}

For all systems, $S$ exhibits a sharp onset, marking the transition from the pinned state to the stick–slip regime. The ordering of the depinning torques $\tau_{\mathrm{depin}}$, is qualitatively consistent with the effective energy barriers identified in Sec.~\ref{subsec:shell_rotation}. 

Conversely, the reverse sweep reveals a pinning transition at a characteristic torque $\tau_{\mathrm{pin}}$. The corresponding depinning and pinning thresholds for all systems are compiled in Table~\ref{tab:depinning}. A comparison of the ratios of the depinning torques in the dynamic analysis and the torque ratios predicted from the static PN-energy landscape is given in Appendix~\ref{sec:depinning_torque}. 

We find that for most systems, $S$ asymptotically converges to unity after $\tau_\text{depin}$ is exceeded. A second increase of $S$ in the $\alpha=1.52$, $N=44$ and $\alpha=3.07$, $N=83$ systems stick-slip regime is attributed to a partial dynamical unlocking of previously more strongly pinned regions of the outer shell, increasing the overall sliding efficiency. 

We observe a noticeable hysteresis in the $\alpha=4.10$, $N=73$ system which coincidentally falls in line with the sliding efficiency of the $N=72$ system. Extended simulations show that, in the down–sweep, the outer shell relaxes into a metastable configuration that differs from the ground state, independent of the phase of the rotation. When initialized at high drive, the system follows a different path in configuration space and settles into a stationary rotational state with reduced corrugation. These lower–barrier states are dynamically accessible on the down–sweep but not when starting from the ground state configuration. We highlighted the differing configurations in Fig.~\ref{fig:stick_slip_friction}(c) and (d) with \textbf{A} and \textbf{B} denoting the different crystal configurations and the arrows indicating the directionality of the torque-sweep. While a more detailed microscopic account of the geometric differences is beyond the scope of this work, our results indicate that the observed hysteresis originates from high–torque–induced rearrangements of the outer shell that reduce the effective angular corrugation and sustain rotation below the static depinning threshold. By contrast, in the other systems the absence of a comparable hysteresis suggests that such metastable states are either not realized or do not remain stable upon adiabatic deceleration.

Closer inspection of the high driving torque regime reveals a small local minimum in $S$ at driving torques between $\tau_\text{drive}/\tau' \approx 25$ and $35$. While this feature is most pronounced in the $\alpha=3.07$ systems, it consistently appears across all investigated configurations and points to a regime of enhanced energy dissipation. 

A possible explanation is that, in this regime, the outer shell rotates at a frequency that aligns unfavorably with the periodicity of the corrugation potential, such that ions repeatedly encounter the flanks of the potential barriers in a dynamically inefficient phase. This leads to an increased energy loss and, consequently, to a reduction in $S$. Extended simulations showed a broadening of the angular velocity distribution within this range of driving torques, which supports the given explanation showing that some ions experience slightly different local barrier conditions and dynamical phases along their trajectory. Although this interpretation is consistent with the observed features, we note that the approximate alignment of the minima across systems with different corrugation periodicities suggests that additional collective or geometric effects may also play a role.

\subsubsection{Inter- and intra-shell friction}
To quantify how uniformly the self-organized outer shell rotates, we compute the average angular velocity of each outer-shell ion and normalize it by the maximum particle value,
\begin{equation}
    \overline{\Omega}_i = \frac{\langle\Omega_i\rangle}{\max(\langle\Omega_i\rangle)}.
\end{equation}
For each system the driving torque is adjusted such that the shell sliding efficiency is $\overline{\Omega}=0.9$, and $\overline{\Omega}_i$ is evaluated in the stationary rotating state. Ions located close to the $z$ poles are omitted due to inconsistent values of $\overline{\Omega}_i$.

Figure~\ref{fig:ang_vel_vs_z} shows $\overline{\Omega}_i$ versus the ions' axial positions $z_i$ for the $\alpha=1.52$ and $\alpha=3.07$ systems. The $\alpha=1.0$ and $\alpha=4.10$ cases are not shown since those systems yield near-flat profiles with $\overline{\Omega}_i \approx 1$ across $z$. Although the corrugation generally varies along $z$, a flat profile indicates that tangential Coulomb coupling within the outer shell redistributes corrugation–induced drag and homogenizes the particle speeds. By contrast, $\alpha=1.52,\,N=44$ and $\alpha=3.07,\,N=83$ show a reduction of $\overline{\Omega}_i$ toward the crystal center, whereas their partner systems ($N=50$ and $N=82$) remain homogeneous in $\overline{\Omega}_i$. Our results show that a change of $N$ at the same $\alpha$ can lead to spatially different coupling strengths to the corrugation potential of the inner shell. When this localized friction dominates over intra-shell coupling, ring-like domains of different angular velocities develop within the outer shell creating an inhomogeneous velocity profile along $z$. 
To verify that these inhomogeneous angular-velocity profiles do not reflect a loss of shell integrity, we additionally analyzed the radial fluctuations of the rotating outer shell in the rotational steady state. We observe that the average radial fluctuations of the ions in the outer shell are more than an order of magnitude smaller than the mean inter-shell spacing. Therefore, shell-crossing events or radial exchange between shells in the investigated regime are not observed and rearrangements remain confined to motion along the shell surface.

More broadly, self-organized systems driven across a static corrugation potential can realize coexisting fast and slow moving regions giving rise to 1D nanofriction that, together with the 2D inter-shell contribution, constitutes a multidimensional friction process. A schematic representation of this concept is shown in Fig.~\ref{fig:multi_dim_friction}. A closer analysis of those inter- and intra-shell friction processes is beyond the scope of the present work and remains an interesting subject for future investigation.

\medskip
\noindent\textbf{Conclusion C} \\
We identified distinct frictional regimes between shell structures in 3D Coulomb crystals, including pinned, stick-slip, and smooth sliding. The depinning torque and sliding efficiency strongly depends on ion number, the geometry of the inner shells and the aspect ratio of the trapping potential $\alpha$. Our findings of $\tau_\text{depin}$ qualitatively match the expectations gained from the effective energy barriers calculated in Sec.~\ref{subsec:shell_rotation}. 
The analysis of the systems dynamics have revealed a plethora of more complex inter-shell friction phenomena such as multidimensional friction, dynamical locking processes and differences in the hysteresis of the sliding velocity for different system sizes and values of $\alpha$. While our findings offer only a first look into the dynamical interaction processes governing inter-shell friction, the observed dynamics point to a rich complexity which remains an interesting subject for future research.

\section{Prospects for Experimental Realization}
\label{subsec:exp_feasibility}
To experimentally observe the predicted ion-number dependence of the depinning threshold for the rotation of the outer shell, we propose preparing two crystals at fixed aspect ratio $\alpha$ that differ in shell commensurability and inner-shell occupancy, e.g. the $N\!=\!72$ and  $N\!=\!73$ systems at $\alpha\!=\!4.10$. These two systems are easily distinguishable on EMCCD images due to the additional ion in the inner shell of the $N=73$ crystal compared to the $N=72$ system. Preparation follows standard Doppler-annealing with repeated image-based state selection until the ground state configuration has been reached. Slightly lifting the radial degeneracy by a few percent ($\omega_x \lesssim \omega_y$) fixes the crystal orientation without significantly altering the inter-shell geometry and therefore the depinning threshold.

A key experimental concern is rf-induced micromotion (MM). In a linear Paul trap, the rf field drives motion at $\Omega_{\mathrm{rf}}$ in the radial $x$–$y$ plane. The local MM vector is aligned with $(x,-y)$, i.e. parallel to the rf quadrupole field. 
Therefore, ions located exactly on one of the radial axes experience purely radial MM, while ions at generic azimuthal positions acquire a finite tangential component which changes its direction in each quadrant so that the net rf drive along $\varphi$ averages to zero over the full shell \cite{berkeland_Micromotion_1998}. Thus, while individual ions do experience tangential MM, this does not by itself trigger the depinning of the outer shell. For our parameters ($^{9}$Be$^+$, $\Omega_{\mathrm{rf}} \approx \SI{35}{\mega\hertz}$, $\omega_r/2\pi \approx \SI{1.4}{\mega\hertz}$, $q_r \approx 0.186$), even the largest MM amplitudes at the outer shell (max. radius of $\sim\SI{36}{\micro\metre}$) remain $\sim 3$–$4~\mu$m, well below the typical inter-ion spacing on the order of $\sim\SI{30}{\micro\metre}$. Nonetheless, rf-induced micromotion enhances the coupling of residual electric-field noise to the secular modes and can heat the crystal \cite{kalincev_MotionalHeating_2021}, so an rf drive with high spectral purity and low noise will be required.

The relevant thermal budget is set by Doppler cooling versus residual heating in the secular modes and can be kept at the $\si{\milli\kelvin}$ level with an axial cooling beam and standard excess-MM compensation. 

Rotation of the outer shell can be triggered by applying a near-resonant push beam, slightly blue detuned from the dipole allowed cooling transition, e.g. $\SI{313}{\nano\metre}$ for Be$^+$. By illuminating the outer rim ions, the push beam exerts a tangential radiation-pressure force on the outer shell. Assuming a perfectly tangential beam, the average radiation–pressure force per illuminated ion is
\begin{equation}
F_\varphi=\hbar k\,\Gamma_{\rm sc}(\Delta,I)
\end{equation}
with the scattering rate
\begin{equation}
\Gamma_{\rm sc}=\tfrac{\Gamma}{2}\,\dfrac{s}{1+s+(2\Delta/\Gamma)^2},
\end{equation}
where $s=I/I_{\rm sat}$ is the saturation parameter, $\Delta$ the laser detuning and $\Gamma$ the natural linewidth. The total torque created by the laser can be calculated by
\begin{equation}
\tau_\text{L}(I)=\sum_{i} r_i \, F_{\varphi}\,,
\end{equation}
with the radial distances $r_i$ of the ions to the $z$-axis.
In the unsaturated limit ($s\ll1$) one has $\Gamma_{\rm sc}\propto I$, so the threshold intensity approximates the ratio of the depinning torques $I_{\rm th}^{(N)}/I_{\rm th}^{(N')} \approx \tau_{\rm depin}^{(N)}/\tau_{\rm depin}^{(N')}$. For further discussion on the ratio of the depinning torques, see Appendix~\ref{sec:depinning_torque}. 

The rotation would then be visible by an azimuthal blurring of the outer-shell ion spots into short arcs or rings for longer exposures on the EMCCD images when the threshold intensity of the beam has been reached. For the $N=72$ vs.\ $N=73$ pair at $\alpha=4.10$, our simulations predict a depinning-torque ratio on the order of $ \sim 4.1$, which should be observable as a comparable ratio of threshold intensities under otherwise identical conditions. Thus, the predicted sensitivity to particle number could be observed experimentally by comparing, at fixed $\alpha$, the threshold push-beam intensity at which the outer-shell fluorescence changes from localized spots to azimuthally blurred arcs, provided the corresponding ground state configurations can be prepared reproducibly.

Finally, to suppress rigid-body rotation of the entire crystal, one may either use a second ion species of lighter mass to induce a pinning impurity (similar to Duca et al. \cite{duca_orientational_2023}) or use differential laser forces on the inner and outer shell (similar to Kiethe et al. \cite{Kiethe_Nanofriction_2018}) to keep the inner shell from rotating. 
In summary, reproducing the exact 3D ground state configurations is experimentally demanding, requiring imaging from two different angles, precise excess-micromotion compensation, and fine control of the secular anisotropy, but it is well within current ion-trapping capabilities and therefore technically feasible.

\section{Summary}
\label{sec:summary}
In this work, we used molecular-dynamics simulations of three-dimensional (3D) ion Coulomb crystals in a linear Paul trap to investigate the formation and scaling of concentric shells and study them as a model system for two-dimensional (2D) nanofriction by rotating the outer shell around the corrugation potential of the inner core. In summary, our three main conclusions are:
\par\medskip
\noindent\textbf{Scaling of shell formation with ion number in finite systems} \\
We have demonstrated that the transition boundaries between different shell counts follow nearly linear trends when mapping the number of shells on the particle number $N$ and the aspect ratio of the trapping potential $\alpha$. Our findings show that the shell count is well captured by a power-law scaling with $\alpha/N$. Since the number of shells can be directly identified in fluorescence images and $\alpha$ is experimentally controlled through the trap parameters, this scaling provides a practical way to estimate the ion number in spheroidal Coulomb crystals from their visible shell structure.
\par\medskip
\noindent\textbf{Collective effects on inter-shell rotation} \\
We calculated a Peierls--Nabarro-type potential for the rotation of the outer shell against the static inner core, by treating the angle of rotation as a collective coordinate. We find that changing the the ion number $N$ by just one can alter the effective rotational barrier by a factor of up to $\sim 7$. In extreme cases, changing $N$ by a few can modify the effective barrier by a factor of up to $60$. This demonstrates that the ion number and the resulting change in crystal configuration acts as a discrete control knob for the rotational barrier in spheroidal Coulomb crystals.

A key finding is that these barrier fluctuations are not simply a consequence of global magic-number stability, as we find no meaningful correlation between the large changes in the effective barrier and the finite-size stability of the ground state. 
We quantified a geometric commensurability measure and show that while inter-shell commensurability represents an important structural factor in the observed fluctuations, it does not universally determine the full barrier. 
Instead, the barrier reflects a system-dependent interplay between the inter-shell interaction, the structural response of the outer shell, and the confining contribution that stabilizes the crystal, providing the restoring environment for relative shell motion. 3D Coulomb crystals therefore provide a model platform for nanofriction in finite systems with curved interfaces, where the frictional response is shaped both by commensurability and by the structural constraints required for mechanical stability.
\par\medskip
\noindent\textbf{Frictional regimes and domain formation in rotating inhomogeneous Coulomb crystals} \\
We identified pinned, stick-slip, and smooth-sliding regimes by applying rotational torques to the outer shells of systems with particularly large differences in the rotational barrier. The observed depinning thresholds qualitatively follow the trends expected from the quasistatic energy barriers, indicating that the Peierls--Nabarro-type landscape captures the relevant frictional resistance. At the same time, the driven dynamics reveal effects that are not visible from the static barrier alone, including torque-induced metastable states, hysteresis, and dynamical unlocking processes.

Finally, we found that nanofriction in these inhomogeneous crystals has a multidimensional character: We observe sliding domains with different rotation speeds along the axis of rotation. These emerge from spatially varying corrugation conditions of the inhomogeneous crystal. The full sliding dynamics are therefore captured by a combination of 2D friction between the shells and 1D shear between neighboring domains.
\par\medskip
\noindent
Taken together, our results show that friction between shells in finite 3D Coulomb crystals is shaped by the corrugation of the inner shell and the relaxation trajectories of outer-shell particles. This makes trapped Coulomb crystals a tunable model system for curved-interface nanofriction. Configurations with a high barrier for outer-shell rotation may be used for stabilizing systems against orientational melting, whereas low-barrier configurations could be exploited for ultra-low-friction nanomechanical systems such as ion-based nanorotors, gyroscopes, or ultrasensitive torque sensors.

\begin{acknowledgments}
We would like to thank Ramil Nigmatullin for providing the basis of the simulation codes that were used during our research.
We acknowledge support by the projects 18SIB05 ROCIT and 20FUN01 TSCAC.
These projects have received funding from the EMPIR programme cofinanced by the Participating States and from the European Union’s Horizon 2020 research and innovation programme.
This project has been funded by the Deutsche Forschungsgemeinschaft (DFG, German Research Foundation) under Germany’s Excellence Strategy -- EXC-2123 QuantumFrontiers–390837967 and through CRC 1227 (DQ-mat), project A07.
\end{acknowledgments}

\bibliography{paper}

@article{drewsen_large_1998,
  author = {Drewsen, M. and Brodersen, C. and Hornek{\ae}r, L. and Hangst, J. S. and Schiffer, J. P.},
  title = {Large {{Ion Crystals}} in a {{Linear Paul Trap}}},
  journal = {Phys. Rev. Lett.},
  volume = {81},
  number = {14},
  pages = {2878--2881},
  year = {1998},
  month = {oct},
  issn = {0031-9007, 1079-7114},
  doi = {10.1103/PhysRevLett.81.2878},
  urldate = {2024-04-25},
  langid = {english},
  copyright = {http://link.aps.org/licenses/aps-default-license},
}

@article{kiethe_probing_2017,
  author = {Kiethe, J. and Nigmatullin, R. and Kalincev, D. and Schmirander, T. and Mehlst{\"a}ubler, T. E.},
  title = {Probing Nanofriction and {{Aubry-type}} Signatures in a Finite Self-Organized System},
  journal = {Nat. Commun.},
  volume = {8},
  number = {1},
  pages = {15364},
  year = {2017},
  month = {may},
  issn = {2041-1723},
  doi = {10.1038/ncomms15364},
  urldate = {2024-05-03},
  langid = {english},
}

@article{long_2D_FK_2010,
  title = {Application of {{Two-Dimensional Frenkel}}--{{Kontorova Model}} to {{Nanotribology}}},
  author = {{Cang-Long}, Wang and {Wen-Shan}, Duan and Yang, Yang and {Jian-Min}, Chen},
  year = 2010,
  month = jul,
  journal = {Commun. Theor. Phys.},
  volume = {54},
  number = {1},
  pages = {112--116},
  issn = {0253-6102},
  doi = {10.1088/0253-6102/54/1/22},
  urldate = {2024-07-23}
}

@article{radzvilavicius_TopologicalDefectMotifs_2011,
  title = {Topological Defect Motifs in Two-Dimensional {{Coulomb}} Clusters},
  author = {Radzvilavi{\v c}ius, A and Anisimovas, E},
  year = 2011,
  month = sep,
  journal = {J. Phys.: Condens. Matter},
  volume = {23},
  number = {38},
  pages = {385301},
  issn = {0953-8984, 1361-648X},
  doi = {10.1088/0953-8984/23/38/385301},
  urldate = {2025-10-29}
}

@article{duca_orientational_2023,
  author = {Duca, Lucia and Mizukami, Naoto and Perego, Elia and Inguscio, Massimo and Sias, Carlo},
  title = {Orientational {{Melting}} in a {{Mesoscopic System}} of {{Charged Particles}}},
  journal = {Phys. Rev. Lett.},
  volume = {131},
  number = {8},
  pages = {083602},
  year = {2023},
  month = {aug},
  issn = {0031-9007, 1079-7114},
  doi = {10.1103/PhysRevLett.131.083602},
  urldate = {2025-10-10},
  langid = {english},
}

@article{Hasse_cylindrical_Coulomb_1990,
  author  = {Hasse, R. W. and Schiffer, J. P.},
  title   = {The structure of the cylindrically confined {Coulomb} lattice},
  year = 1990,
  month = nov,
  journal = {Ann. Phys.},
  volume = {203},
  number = {2},
  pages = {419--448},
  publisher = {Elsevier BV},
  issn = {0003-4916},
  doi = {10.1016/0003-4916(90)90177-p},
  urldate = {2024-09-17},
  copyright = {https://www.elsevier.com/tdm/userlicense/1.0/},
  langid = {english}
}

@article{solis_annealing_2014,
  author = {Juan Frausto-Sol{\'\i}s and Ernesto Li{\~n}{\'a}n-Garc{\'\i}a and Guillermo Santamar{\'\i}a-Bonfil},
  title = {Tuned Simulated Annealing Based on {Boltzmann} and {Bose}--{Einstein} Distribution Applied to {MAXSAT} Problem},
  journal = {J. Asian Sci. Res.},
  volume = {4},
  number = {1},
  pages = {14--26},
  year = {2014},
  url = {https://archive.aessweb.com/index.php/5003/article/download/3594/5700},
}

@article{Rahman_OCP_1986,
  author = {Rahman, A. and Schiffer, J. P.},
  title = {Structure of a {{One-Component Plasma}} in an {{External Field}}: {{A Molecular-Dynamics Study}} of {{Particle Arrangement}} in a {{Heavy-Ion Storage Ring}}},
  journal = {Phys. Rev. Lett.},
  volume = {57},
  number = {9},
  pages = {1133--1136},
  year = {1986},
  month = {sep},
  issn = {0031-9007},
  doi = {10.1103/PhysRevLett.57.1133},
  urldate = {2024-12-10},
  langid = {english},
  shorttitle = {Structure of a {{One-Component Plasma}} in an {{External Field}}},
  copyright = {http://link.aps.org/licenses/aps-default-license},
}

@incollection{Totsuji_plasma_physics_1987,
  author = {Totsuji, Hiroo},
  title = {Static and {{Dynamic Properties}} of {{Strongly-Coupled Classical One-Component Plasmas}}: {{Numerical Experiments}} on {{Supercooled Liquid State}} and {{Simulation}} of {{Ion Plasma}} in the {{Penning Trap}}},
  booktitle = {Strongly {{Coupled Plasma Physics}}},
  editor = {Rogers, Forrest J. and Dewitt, Hugh E.},
  pages = {19--33},
  year = {1987},
  publisher = {Springer US},
  address = {Boston, MA},
  isbn = {978-1-4612-9053-7 978-1-4613-1891-0},
  doi = {10.1007/978-1-4613-1891-0_2},
  urldate = {2024-12-10},
  langid = {english},
  shorttitle = {Static and {{Dynamic Properties}} of {{Strongly-Coupled Classical One-Component Plasmas}}},
}

@article{Dubin_computer_simulation_IC_1988,
  author = {Dubin, D. and O'Neil, T.},
  title = {Computer Simulation of Ion Clouds in a {{Penning}} Trap},
  journal = {Phys. Rev. Lett.},
  volume = {60},
  number = {6},
  pages = {511--514},
  year = {1988},
  month = {feb},
  issn = {0031-9007},
  doi = {10.1103/PhysRevLett.60.511},
  urldate = {2024-12-10},
  langid = {english},
}

@article{Okada_characterization_2010,
  author = {Okada, K. and Wada, M. and Takayanagi, T. and Ohtani, S. and Schuessler, H. A.},
  title = {Characterization of Ion {{Coulomb}} Crystals in a Linear {{Paul}} Trap},
  journal = {Phys. Rev. A},
  volume = {81},
  number = {1},
  pages = {013420},
  year = {2010},
  month = {jan},
  issn = {1050-2947, 1094-1622},
  doi = {10.1103/PhysRevA.81.013420},
  urldate = {2024-12-10},
  langid = {english},
  copyright = {http://link.aps.org/licenses/aps-default-license},
}

@article{Birkl_multi_shell_ring_1992,
  author = {Birkl, G. and Kassner, S. and Walther, H.},
  title = {Multiple-Shell Structures of Laser-Cooled {{$^{24}$Mg}}$^+$ Ions in a Quadrupole Storage Ring},
  journal = {Nature},
  volume = {357},
  number = {6376},
  pages = {310--313},
  year = {1992},
  month = {may},
  issn = {0028-0836, 1476-4687},
  doi = {10.1038/357310a0},
  urldate = {2024-12-10},
  langid = {english},
  copyright = {http://www.springer.com/tdm},
}

@article{Holmberg_Friction_2017,
  author = {Holmberg, Kenneth and Erdemir, Ali},
  title = {Influence of Tribology on Global Energy Consumption, Costs and Emissions},
  journal = {Friction},
  volume = {5},
  number = {3},
  pages = {263--284},
  year = {2017},
  month = {sep},
  issn = {2223-7690, 2223-7704},
  doi = {10.1007/s40544-017-0183-5},
  urldate = {2025-01-13},
  langid = {english},
}

@article{Paul1990,
  author = {Paul, Wolfgang},
  title = {Electromagnetic traps for charged and neutral particles},
  journal = {Rev. Mod. Phys.},
  volume = {62},
  issue = {3},
  pages = {531--540},
  year = {1990},
  month = {Jul},
  publisher = {American Physical Society},
  doi = {10.1103/RevModPhys.62.531},
  numpages = {0},
}

@article{Han_Superlubricity_2020,
  author = {Han, Juan-Fang and Liu, Bo and Duan, Wen-Shan},
  title = {The Superlubricity of the Special Material with Hexagonal Symmetry in a Two-Dimensional {{Frenkel}}--{{Kontorova}} Model},
  journal = {Indian J. Phys.},
  volume = {94},
  number = {4},
  pages = {521--526},
  year = {2020},
  month = {apr},
  issn = {0973-1458, 0974-9845},
  doi = {10.1007/s12648-019-01498-4},
  urldate = {2025-10-06},
  langid = {english},
}

@article{FrenkelKontorova1938,
  author = {Frenkel, Yakov and Kontorova, Tatiana},
  title = {On the theory of plastic deformation and twinning},
  journal = {Zh. Eksp. Teor. Fiz.},
  volume = {8},
  pages = {1340--1348},
  year = {1938},
  note = {In Russian},
}

@article{Prandtl1928,
  author = {Prandtl, L.},
  title = {Ein {{Gedankenmodell}} Zur Kinetischen {{Theorie}} Der Festen {{K{\"o}rper}}},
  journal = {Z. Angew. Math. Mech.},
  volume = {8},
  number = {2},
  pages = {85--106},
  year = {1928},
  month = {jan},
  issn = {0044-2267, 1521-4001},
  doi = {10.1002/zamm.19280080202},
  urldate = {2025-10-10},
  langid = {english},
  copyright = {http://onlinelibrary.wiley.com/termsAndConditions\#vor},
}

@article{Tomlinson1929,
  author = {Tomlinson, G. A.},
  title = {A molecular theory of friction},
  journal = {Philos. Mag. Ser. 7},
  volume = {7},
  number = {46},
  pages = {905--939},
  year = {1929},
}

@article{Drewsen_ICCs_2015,
  author = {Drewsen, Michael},
  title = {Ion {{Coulomb}} Crystals},
  journal = {Physica B},
  volume = {460},
  pages = {105--113},
  year = {2015},
  month = {mar},
  issn = {09214526},
  doi = {10.1016/j.physb.2014.11.050},
  urldate = {2024-11-15},
  langid = {english},
}

@article{Dubin_phase_transition_1993,
  author = {Dubin, Daniel H. E.},
  title = {Theory of Structural Phase Transitions in a Trapped {{Coulomb}} Crystal},
  journal = {Phys. Rev. Lett.},
  volume = {71},
  number = {17},
  pages = {2753--2756},
  year = {1993},
  month = {oct},
  issn = {0031-9007},
  doi = {10.1103/PhysRevLett.71.2753},
  urldate = {2025-03-04},
  langid = {english},
  copyright = {http://link.aps.org/licenses/aps-default-license},
}

@article{Ruffert_Domain_2024,
  author = {R{\"u}ffert, L.-A. and Dijck, E. A. and Timm, L. and {L{\'o}pez-Urrutia}, J. R. Crespo and Mehlst{\"a}ubler, T. E.},
  title = {Domain Formation and Structural Stabilities in Mixed-Species {{Coulomb}} Crystals Induced by Sympathetically Cooled Highly Charged Ions},
  journal = {Phys. Rev. A},
  volume = {110},
  number = {6},
  pages = {063110},
  year = {2024},
  month = {dec},
  issn = {2469-9926, 2469-9934},
  doi = {10.1103/PhysRevA.110.063110},
  urldate = {2025-03-04},
  langid = {english},
}

@article{Kiethe_Nanofriction_2018,
  author = {Kiethe, J and Nigmatullin, R and Schmirander, T and Kalincev, D and Mehlst{\"a}ubler, T E},
  title = {Nanofriction and Motion of Topological Defects in Self-Organized Ion {{Coulomb}} Crystals},
  journal = {New J. Phys.},
  volume = {20},
  number = {12},
  pages = {123017},
  year = {2018},
  month = {dec},
  publisher = {IOP Publishing},
  issn = {1367-2630},
  doi = {10.1088/1367-2630/aaf3d5},
  urldate = {2025-03-19},
  langid = {english},
}

@article{benassi_nanofriction_2011,
  author = {Benassi, A. and Vanossi, A. and Tosatti, E.},
  title = {Nanofriction in Cold Ion Traps},
  journal = {Nat. Commun.},
  volume = {2},
  number = {1},
  pages = {236},
  year = {2011},
  month = {mar},
  issn = {2041-1723},
  doi = {10.1038/ncomms1230},
  urldate = {2025-01-09},
  langid = {english},
}

@article{braiman_Symmetrybreaking_1990,
  author = {Braiman, Y. and Baumgarten, J. and Jortner, Joshua and Klafter, J.},
  title = {Symmetry-Breaking Transition in Finite {{Frenkel-Kontorova}} Chains},
  journal = {Phys. Rev. Lett.},
  volume = {65},
  number = {19},
  pages = {2398--2401},
  year = {1990},
  month = {nov},
  publisher = {American Physical Society},
  doi = {10.1103/PhysRevLett.65.2398},
  urldate = {2025-03-20},
}

@article{mandelli_Nanofriction_2013,
  author = {Mandelli, D. and Vanossi, A. and Tosatti, E.},
  title = {Stick-Slip Nanofriction in Trapped Cold Ion Chains},
  journal = {Phys. Rev. B},
  volume = {87},
  number = {19},
  pages = {195418},
  year = {2013},
  month = {may},
  publisher = {American Physical Society},
  doi = {10.1103/PhysRevB.87.195418},
  urldate = {2025-03-20},
}

@article{bylinskii_friction_2015,
  author = {Bylinskii, Alexei and Gangloff, Dorian and Vuleti{\'c}, Vladan},
  title = {Tuning Friction Atom-by-Atom in an Ion-Crystal Simulator},
  journal = {Science},
  volume = {348},
  number = {6239},
  pages = {1115--1118},
  year = {2015},
  month = {jun},
  issn = {0036-8075, 1095-9203},
  doi = {10.1126/science.1261422},
  urldate = {2025-10-10},
  langid = {english},
  copyright = {http://www.sciencemag.org/about/science-licenses-journal-article-reuse},
}

@article{Mate_AtomicscaleFrictionTungsten_1987,
  author = {Mate, C. Mathew and McClelland, Gary M. and Erlandsson, Ragnar and Chiang, Shirley},
  title = {Atomic-Scale Friction of a Tungsten Tip on a Graphite Surface},
  journal = {Phys. Rev. Lett.},
  volume = {59},
  number = {17},
  pages = {1942--1945},
  year = {1987},
  month = {oct},
  issn = {0031-9007},
  doi = {10.1103/PhysRevLett.59.1942},
  urldate = {2025-05-06},
  langid = {english},
  copyright = {http://link.aps.org/licenses/aps-default-license},
}

@article{Andersson_Friction_2020,
  author = {Andersson, David and De Wijn, Astrid S.},
  title = {Understanding the Friction of Atomically Thin Layered Materials},
  journal = {Nat. Commun.},
  volume = {11},
  number = {1},
  pages = {420},
  year = {2020},
  month = {jan},
  issn = {2041-1723},
  doi = {10.1038/s41467-019-14239-2},
  urldate = {2025-01-09},
  langid = {english},
}

@article{dienwiebel_SuperlubricityGraphite_2004,
  author = {Dienwiebel, Martin and Verhoeven, Gertjan S. and Pradeep, Namboodiri and Frenken, Joost W. M. and Heimberg, Jennifer A. and Zandbergen, Henny W.},
  title = {Superlubricity of {{Graphite}}},
  journal = {Phys. Rev. Lett.},
  volume = {92},
  number = {12},
  pages = {126101},
  year = {2004},
  month = {mar},
  issn = {0031-9007, 1079-7114},
  doi = {10.1103/PhysRevLett.92.126101},
  urldate = {2025-05-06},
  langid = {english},
  copyright = {http://link.aps.org/licenses/aps-default-license},
}

@article{xu_VanishingStickSlip_2011,
  author = {Xu, Liang and Ma, Tian-Bao and Hu, Yuan-Zhong and Wang, Hui},
  title = {Vanishing Stick--Slip Friction in Few-Layer Graphenes: The Thickness Effect},
  journal = {Nanotechnology},
  volume = {22},
  number = {28},
  pages = {285708},
  year = {2011},
  month = {jul},
  issn = {0957-4484, 1361-6528},
  doi = {10.1088/0957-4484/22/28/285708},
  urldate = {2025-02-12},
  shorttitle = {Vanishing Stick--Slip Friction in Few-Layer Graphenes},
}

@article{voisin_LongTermFriction_2007,
  title = {Long Term Friction: {{From}} Stick-slip to Stable Sliding},
  shorttitle = {Long Term Friction},
  author = {Voisin, Christophe and Renard, Fran{\c c}ois and Grasso, Jean-Robert},
  year = 2007,
  month = jul,
  journal = {Geophys. Res. Lett.},
  volume = {34},
  number = {13},
  pages = {},
  issn = {0094-8276, 1944-8007},
  doi = {10.1029/2007GL029715},
  urldate = {2025-10-29},
  copyright = {http://onlinelibrary.wiley.com/termsAndConditions\#vor},
  langid = {english}
}

@article{gourdon_TransitionsSmoothComplex_2003,
  author = {Gourdon, Delphine and Israelachvili, Jacob N.},
  title = {Transitions between Smooth and Complex Stick-Slip Sliding of Surfaces},
  journal = {Phys. Rev. E},
  volume = {68},
  number = {2},
  pages = {021602},
  year = {2003},
  month = {aug},
  issn = {1063-651X, 1095-3787},
  doi = {10.1103/PhysRevE.68.021602},
  urldate = {2025-05-08},
  langid = {english},
  copyright = {http://link.aps.org/licenses/aps-default-license},
}

@article{drummond_DynamicPhaseTransitions_2001,
  author = {Drummond, Carlos and Israelachvili, Jacob},
  title = {Dynamic Phase Transitions in Confined Lubricant Fluids under Shear},
  journal = {Phys. Rev. E},
  volume = {63},
  number = {4},
  pages = {041506},
  year = {2001},
  month = {mar},
  issn = {1063-651X, 1095-3787},
  doi = {10.1103/PhysRevE.63.041506},
  urldate = {2025-05-08},
  langid = {english},
  copyright = {http://link.aps.org/licenses/aps-default-license},
}

@article{brazda_Aubry_2018,
  author = {Brazda, T. and Silva, A. and Manini, N. and Vanossi, A. and Guerra, R. and Tosatti, E. and Bechinger, C.},
  title = {Experimental {{Observation}} of the {{Aubry Transition}} in {{Two-Dimensional Colloidal Monolayers}}},
  journal = {Phys. Rev. X},
  volume = {8},
  number = {1},
  pages = {011050},
  year = {2018},
  month = {mar},
  issn = {2160-3308},
  doi = {10.1103/PhysRevX.8.011050},
  urldate = {2025-05-08},
  langid = {english},
}

@incollection{aubry_ConceptTransitions_1978,
  author = {Aubry, Serge},
  title = {The {{New Concept}} of {{Transitions}} by {{Breaking}} of {{Analyticity}} in a {{Crystallographic Model}}},
  booktitle = {Solitons and {{Condensed Matter Physics}}},
  editor = {Cardona, Manuel and Fulde, Peter and Queisser, Hans-Joachim and Bishop, Alan R. and Schneider, Toni},
  volume = {8},
  pages = {264--277},
  year = {1978},
  publisher = {Springer Berlin Heidelberg},
  address = {Berlin, Heidelberg},
  isbn = {978-3-642-81293-4 978-3-642-81291-0},
  doi = {10.1007/978-3-642-81291-0_28},
  urldate = {2025-05-08},
}

@article{peyrard_CriticalTransition_1983,
  author = {Peyrard, M and Aubry, S},
  title = {Critical Behaviour at the Transition by Breaking of Analyticity in the Discrete {{Frenkel-Kontorova}} Model},
  journal = {J. Phys. C: Solid State Phys.},
  volume = {16},
  number = {9},
  pages = {1593--1608},
  year = {1983},
  month = {mar},
  issn = {0022-3719},
  doi = {10.1088/0022-3719/16/9/005},
  urldate = {2025-05-08},
}

@article{binnig_AtomicForceMicroscope_1986,
  author = {Binnig, G. and Quate, C. F. and Gerber, {\relax Ch}.},
  title = {Atomic {{Force Microscope}}},
  journal = {Phys. Rev. Lett.},
  volume = {56},
  number = {9},
  pages = {930--933},
  year = {1986},
  month = {mar},
  issn = {0031-9007},
  doi = {10.1103/PhysRevLett.56.930},
  urldate = {2025-05-12},
  langid = {english},
  copyright = {http://link.aps.org/licenses/aps-default-license},
}

@article{martin_AtomicForceMicroscope_1987,
  author = {Martin, Y. and Williams, C. C. and Wickramasinghe, H. K.},
  title = {Atomic Force Microscope--Force Mapping and Profiling on a Sub 100-{{{\AA}}} Scale},
  journal = {J. Appl. Phys.},
  volume = {61},
  number = {10},
  pages = {4723--4729},
  year = {1987},
  month = {may},
  issn = {0021-8979, 1089-7550},
  doi = {10.1063/1.338807},
  urldate = {2025-05-12},
  langid = {english},
}

@article{monroe_ScalingIonTrap_2013,
  author = {Monroe, C. and Kim, J.},
  title = {Scaling the {{Ion Trap Quantum Processor}}},
  journal = {Science},
  volume = {339},
  number = {6124},
  pages = {1164--1169},
  year = {2013},
  month = {mar},
  issn = {0036-8075, 1095-9203},
  doi = {10.1126/science.1231298},
  urldate = {2025-05-12},
  langid = {english},
}

@article{cirac_QuantumComputationsCold_1995,
  author = {Cirac, J. I. and Zoller, P.},
  title = {Quantum {{Computations}} with {{Cold Trapped Ions}}},
  journal = {Phys. Rev. Lett.},
  volume = {74},
  number = {20},
  pages = {4091--4094},
  year = {1995},
  month = {may},
  issn = {0031-9007, 1079-7114},
  doi = {10.1103/PhysRevLett.74.4091},
  urldate = {2025-05-12},
  langid = {english},
  copyright = {http://link.aps.org/licenses/aps-default-license},
}

@article{blatt_EntangledStatesTrapped_2008,
  author = {Blatt, Rainer and Wineland, David},
  title = {Entangled States of Trapped Atomic Ions},
  journal = {Nature},
  volume = {453},
  number = {7198},
  pages = {1008--1015},
  year = {2008},
  month = {jun},
  issn = {0028-0836, 1476-4687},
  doi = {10.1038/nature07125},
  urldate = {2025-05-12},
  langid = {english},
  copyright = {http://www.springer.com/tdm},
}

@article{keller_EvaluationTrapinducedSystematic_2016,
  author = {Keller, J and Burgermeister, T and Kalincev, D and Kiethe, J and Mehlst{\"a}ubler, T E},
  title = {Evaluation of Trap-Induced Systematic Frequency Shifts for a Multi-Ion Optical Clock at the 10{\textsuperscript{-19}} Level},
  journal = {J. Phys. Conf. Ser.},
  volume = {723},
  pages = {012027},
  year = {2016},
  month = {jun},
  issn = {1742-6588, 1742-6596},
  doi = {10.1088/1742-6596/723/1/012027},
  urldate = {2025-05-12},
  copyright = {http://iopscience.iop.org/info/page/text-and-data-mining},
}

@article{islam_OnsetQuantumPhase_2011,
  author = {Islam, R. and Edwards, E.E. and Kim, K. and Korenblit, S. and Noh, C. and Carmichael, H. and Lin, G.-D. and Duan, L.-M. and Joseph Wang, C.-C. and Freericks, J.K. and Monroe, C.},
  title = {Onset of a Quantum Phase Transition with a Trapped Ion Quantum Simulator},
  journal = {Nat. Commun.},
  volume = {2},
  number = {1},
  pages = {377},
  year = {2011},
  month = {jul},
  issn = {2041-1723},
  doi = {10.1038/ncomms1374},
  urldate = {2025-05-12},
  langid = {english},
}

@article{britton_EngineeredTwodimensionalIsing_2012,
  author = {Britton, Joseph W. and Sawyer, Brian C. and Keith, Adam C. and Wang, C.-C. Joseph and Freericks, James K. and Uys, Hermann and Biercuk, Michael J. and Bollinger, John J.},
  title = {Engineered Two-Dimensional {{Ising}} Interactions in a Trapped-Ion Quantum Simulator with Hundreds of Spins},
  journal = {Nature},
  volume = {484},
  number = {7395},
  pages = {489--492},
  year = {2012},
  month = {apr},
  issn = {0028-0836, 1476-4687},
  doi = {10.1038/nature10981},
  urldate = {2025-05-12},
  langid = {english},
  copyright = {http://www.springer.com/tdm},
}

@article{bohnet_QuantumSpinDynamics_2016,
  author = {Bohnet, Justin G. and Sawyer, Brian C. and Britton, Joseph W. and Wall, Michael L. and Rey, Ana Maria and {Foss-Feig}, Michael and Bollinger, John J.},
  title = {Quantum Spin Dynamics and Entanglement Generation with Hundreds of Trapped Ions},
  journal = {Science},
  volume = {352},
  number = {6291},
  pages = {1297--1301},
  year = {2016},
  month = {jun},
  issn = {0036-8075, 1095-9203},
  doi = {10.1126/science.aad9958},
  urldate = {2025-05-12},
  langid = {english},
}

@article{gilbert_ShellStructurePhaseMagnetically_1988,
  author = {Gilbert, S. L. and Bollinger, J. J. and Wineland, D. J.},
  title = {Shell-{{Structure Phase}} of {{Magnetically Confined Strongly Coupled Plasmas}}},
  journal = {Phys. Rev. Lett.},
  volume = {60},
  number = {20},
  pages = {2022--2025},
  year = {1988},
  month = {may},
  issn = {0031-9007},
  doi = {10.1103/PhysRevLett.60.2022},
  urldate = {2024-12-10},
  langid = {english},
  copyright = {http://link.aps.org/licenses/aps-default-license},
}

@article{hasse_StructureMadelungEnergy_1991,
  author = {Hasse, R. W. and Avilov, V. V.},
  title = {Structure and {{Madelung}} Energy of Spherical {{Coulomb}} Crystals},
  journal = {Phys. Rev. A},
  volume = {44},
  number = {7},
  pages = {4506--4515},
  year = {1991},
  month = {oct},
  issn = {1050-2947, 1094-1622},
  doi = {10.1103/PhysRevA.44.4506},
  urldate = {2024-12-12},
  langid = {english},
  copyright = {http://link.aps.org/licenses/aps-default-license},
}

@article{ahn_UltrasensitiveTorqueDetection_2020,
  author = {Ahn, Jonghoon and Xu, Zhujing and Bang, Jaehoon and Ju, Peng and Gao, Xingyu and Li, Tongcang},
  title = {Ultrasensitive Torque Detection with an Optically Levitated Nanorotor},
  journal = {Nat. Nanotechnol.},
  volume = {15},
  number = {2},
  pages = {89--93},
  year = {2020},
  month = {feb},
  issn = {1748-3387, 1748-3395},
  doi = {10.1038/s41565-019-0605-9},
  urldate = {2025-05-12},
  langid = {english},
}

@article{shao_MolecularRotorsDesigned_2020,
  author = {Shao, Jian and Zhu, Wenpeng and Zhang, Xiaoyue and Zheng, Yue},
  title = {Molecular Rotors with Designed Polar Rotating Groups Possess Mechanics-Controllable Wide-Range Rotational Speed},
  journal = {npj Comput. Mater.},
  volume = {6},
  number = {1},
  pages = {185},
  year = {2020},
  month = {dec},
  issn = {2057-3960},
  doi = {10.1038/s41524-020-00457-6},
  urldate = {2025-05-12},
  langid = {english},
}

@article{singhania_AccountsAppliedMolecular_2023,
  title = {Accounts of Applied Molecular Rotors and Rotary Motors: Recent Advances},
  shorttitle = {Accounts of Applied Molecular Rotors and Rotary Motors},
  author = {Singhania, Anup and Kalita, Sudeshna and Chettri, Prerna and Ghosh, Subrata},
  year = 2023,
  journal = {Nanoscale Adv.},
  volume = {5},
  number = {12},
  pages = {3177--3208},
  issn = {2516-0230},
  doi = {10.1039/D3NA00010A},
  urldate = {2025-05-12},
  langid = {english}
}

@article{bonitz_SingleelectronControlWigner_2002,
  author = {Bonitz, M and Golubnychiy, V and Filinov, A.V and Lozovik, Yu.E},
  title = {Single-Electron Control of {{Wigner}} Crystallization},
  journal = {Microelectron. Eng.},
  volume = {63},
  number = {1-3},
  pages = {141--145},
  year = {2002},
  month = {aug},
  issn = {01679317},
  doi = {10.1016/S0167-9317(02)00624-X},
  urldate = {2025-05-14},
  langid = {english},
  copyright = {https://www.elsevier.com/tdm/userlicense/1.0/},
}

@article{golubnychiy_ControllingIntershellRotations_2003,
  author = {Golubnychiy, V and Ludwig, P and Filinov, A.V and Bonitz, M},
  title = {Controlling Intershell Rotations in Mesoscopic Electron Clusters},
  journal = {Superlattices Microstruct.},
  volume = {34},
  number = {3-6},
  pages = {219--224},
  year = {2003},
  month = {sep},
  issn = {07496036},
  doi = {10.1016/j.spmi.2004.03.011},
  urldate = {2025-05-14},
  langid = {english},
  copyright = {https://www.elsevier.com/tdm/userlicense/1.0/},
}

@article{berkeland_Micromotion_1998,
  author = {Berkeland, D. J. and Miller, J. D. and Bergquist, J. C. and Itano, W. M. and Wineland, D. J.},
  title = {Minimization of Ion Micromotion in a {{Paul}} Trap},
  journal = {J. Appl. Phys.},
  volume = {83},
  number = {10},
  pages = {5025--5033},
  year = {1998},
  month = {may},
  issn = {0021-8979, 1089-7550},
  doi = {10.1063/1.367318},
  urldate = {2025-05-14},
  langid = {english},
}

@article{kirkSimulatedAnnealing1983,
  author = {Kirkpatrick, S. and Gelatt, C. D. and Vecchi, M. P.},
  title = {Optimization by {{Simulated Annealing}}},
  journal = {Science},
  volume = {220},
  number = {4598},
  pages = {671--680},
  year = {1983},
  month = {may},
  issn = {0036-8075, 1095-9203},
  doi = {10.1126/science.220.4598.671},
  urldate = {2025-06-19},
  langid = {english},
}

@misc{caracciolo_SA_2023,
  author = {Caracciolo, Sergio and Hartmann, Alexander K. and Kirkpatrick, Scott and Weigel, Martin},
  title = {Simulated Annealing, Optimization, Searching for Ground States},
  year = {2023},
  publisher = {arXiv},
  doi = {10.48550/ARXIV.2301.00683},
  urldate = {2024-11-15},
  copyright = {arXiv.org perpetual, non-exclusive license},
}

@article{ohira_2020,
  author = {Ohira, Ryutaro and Mukaiyama, Takashi and Toyoda, Kenji},
  title = {Breaking Rotational Symmetry in a Trapped-Ion Quantum Tunneling Rotor},
  journal = {Phys. Rev. A},
  volume = {101},
  number = {2},
  pages = {022106},
  year = {2020},
  month = {feb},
  issn = {2469-9926, 2469-9934},
  doi = {10.1103/PhysRevA.101.022106},
  urldate = {2025-08-15},
  langid = {english},
}

@article{vuletic_2020,
  author = {Gangloff, Dorian A. and Bylinskii, Alexei and Vuleti{\'c}, Vladan},
  title = {Kinks and Nanofriction: {{Structural}} Phases in Few-Atom Chains},
  journal = {Phys. Rev. Res.},
  volume = {2},
  number = {1},
  pages = {013380},
  year = {2020},
  month = {mar},
  issn = {2643-1564},
  doi = {10.1103/PhysRevResearch.2.013380},
  urldate = {2025-08-18},
  langid = {english},
  shorttitle = {Kinks and Nanofriction},
}

@article{pyka_TopologicalDefectFormation_2013a,
  author = {Pyka, K. and Keller, J. and Partner, H. L. and Nigmatullin, R. and Burgermeister, T. and Meier, D. M. and Kuhlmann, K. and Retzker, A and Plenio, M. B. and Zurek, W. H. and Del Campo, A. and Mehlst{\"a}ubler, T. E.},
  title = {Topological Defect Formation and Spontaneous Symmetry Breaking in Ion {{Coulomb}} Crystals},
  journal = {Nat. Commun.},
  volume = {4},
  number = {1},
  pages = {2291},
  year = {2013},
  month = {aug},
  issn = {2041-1723},
  doi = {10.1038/ncomms3291},
  urldate = {2025-09-22},
  langid = {english},
}

@article{arnold_ProspectsAtomicClocks_2015,
  author = {Arnold, Kyle and Hajiyev, Elnur and Paez, Eduardo and Lee, Chern Hui and Barrett, M. D. and Bollinger, John},
  title = {Prospects for Atomic Clocks Based on Large Ion Crystals},
  journal = {Phys. Rev. A},
  volume = {92},
  number = {3},
  pages = {032108},
  year = {2015},
  month = {sep},
  issn = {1050-2947, 1094-1622},
  doi = {10.1103/PhysRevA.92.032108},
  urldate = {2025-10-01},
  langid = {english},
  copyright = {http://link.aps.org/licenses/aps-default-license},
}

@article{keller_MultiionSpectros_2024,
  author = {Keller, J. and Hausser, H. N. and Richter, I. M. and Nordmann, T. and Bhatt, N. M. and Kiethe, J. and Liu, H. and Benkler, E. and Lipphardt, B. and D{\"o}rscher, S. and Stahl, K. and Klose, J. and Lisdat, C. and Filzinger, M. and Huntemann, N. and Peik, E. and Mehlst{\"a}ubler, T. E.},
  title = {High-Accuracy Multi-Ion Spectroscopy with Mixed-Species {{Coulomb}} Crystals},
  journal = {J. Phys. Conf. Ser.},
  volume = {2889},
  number = {1},
  pages = {012050},
  year = {2024},
  month = {nov},
  issn = {1742-6588, 1742-6596},
  doi = {10.1088/1742-6596/2889/1/012050},
  urldate = {2025-10-01},
}

@article{dubin_Plasmas_1999,
  author = {Dubin, Daniel H. E. and O'Neil, T. M.},
  title = {Trapped Nonneutral Plasmas, Liquids, and Crystals (the Thermal Equilibrium States)},
  journal = {Rev. Mod. Phys.},
  volume = {71},
  number = {1},
  pages = {87--172},
  year = {1999},
  month = {jan},
  issn = {0034-6861, 1539-0756},
  doi = {10.1103/RevModPhys.71.87},
  urldate = {2025-10-06},
  langid = {english},
  copyright = {http://link.aps.org/licenses/aps-default-license},
}

@article{kiesenhofer_TwoDim_2023,
  author = {Kiesenhofer, Dominik and Hainzer, Helene and Zhdanov, Artem and Holz, Philip C. and Bock, Matthias and Ollikainen, Tuomas and Roos, Christian F.},
  title = {Controlling {{two-dimensional Coulomb crystals}} of {{more than}} 100 {{ions}} in a {{monolithic radio-frequency trap}}},
  journal = {PRX Quantum},
  volume = {4},
  number = {2},
  pages = {020317},
  year = {2023},
  month = {apr},
  issn = {2691-3399},
  doi = {10.1103/PRXQuantum.4.020317},
  urldate = {2025-10-23},
  langid = {english},
}

@article{schmidtSpectroscopyUsingQuantum2005,
  author = {Schmidt, P. O. and Rosenband, T. and Langer, C. and Itano, W. M. and Bergquist, J. C. and Wineland, D. J.},
  title = {Spectroscopy {{Using Quantum Logic}}},
  journal = {Science},
  volume = {309},
  number = {5735},
  pages = {749--752},
  year = {2005},
  month = {jul},
  issn = {0036-8075, 1095-9203},
  doi = {10.1126/science.1114375},
  urldate = {2025-10-23},
  langid = {english},
}

@article{kalincev_MotionalHeating_2021,
  title = {Motional Heating of Spatially Extended Ion Crystals},
  author = {Kalincev, D and Dreissen, L S and Kulosa, A P and Yeh, C-H and F{\"u}rst, H A and Mehlst{\"a}ubler, T E},
  year = 2021,
  month = jul,
  journal = {Quantum Sci. Technol.},
  volume = {6},
  number = {3},
  pages = {034003},
  issn = {2058-9565},
  doi = {10.1088/2058-9565/abee99},
  urldate = {2025-11-21}
}

@article{champenoisIonRingLinear2010,
  title = {Ion Ring in a Linear Multipole Trap for Optical Frequency Metrology},
  author = {Champenois, C. and Marciante, M. and {Pedregosa-Gutierrez}, J. and Houssin, M. and Knoop, M. and Kajita, M.},
  year = 2010,
  month = apr,
  journal = {Phys. Rev. A},
  volume = {81},
  number = {4},
  pages = {043410},
  issn = {1050-2947, 1094-1622},
  doi = {10.1103/PhysRevA.81.043410},
  urldate = {2025-12-03},
  copyright = {http://link.aps.org/licenses/aps-default-license},
  langid = {english}
}

@article{baldovin_StronglyCoupled_2024,
  title = {Self-Diffusion in a Strongly Coupled Non-Neutral Plasma},
  author = {Baldovin, Marco and Vallet, Gr{\'e}goire and Hagel, Ga{\"e}tan and Trizac, Emmanuel and Champenois, Caroline},
  year = 2024,
  month = apr,
  journal = {Phys. Rev. A},
  volume = {109},
  number = {4},
  pages = {043116},
  issn = {2469-9926, 2469-9934},
  doi = {10.1103/PhysRevA.109.043116},
  urldate = {2025-12-03},
  langid = {english}
}

@article{Zaris_simulations_2025, 
title={Numerical simulations of three-dimensional ion crystal dynamics in a Penning trap using the fast multipole method}, 
volume={91}, 
DOI={10.1017/S0022377824000990}, 
number={2}, 
journal={J. Plasma Phys.}, 
author={Zaris, John and Johnson, Wes and Shankar, Athreya and Bollinger, John J. and Parker, Scott E.}, 
year={2025}, 
pages={E53}
}

@article{poindron2021thermal,
  title = {Thermal bistability in laser-cooled trapped ions},
  author = {Poindron, Adrien and Pedregosa-Gutierrez, Jofre and Champenois, Caroline},
  journal = {Phys. Rev. A},
  volume = {104},
  issue = {4},
  pages = {043116},
  numpages = {10},
  year = {2021},
  month = {Oct},
  publisher = {American Physical Society},
  doi = {10.1103/PhysRevA.104.043116},
  url = {https://link.aps.org/doi/10.1103/PhysRevA.104.043116}
}

@article{vanossiStaticDynamicFriction2012,
  title = {Static and Dynamic Friction in Sliding Colloidal Monolayers},
  author = {Vanossi, Andrea and Manini, Nicola and Tosatti, Erio},
  year = 2012,
  month = oct,
  journal = {Proc. Natl. Acad. Sci. U.S.A.},
  volume = {109},
  number = {41},
  pages = {16429--16433},
  issn = {0027-8424, 1091-6490},
  doi = {10.1073/pnas.1213930109},
  urldate = {2025-01-10},
  langid = {english}
}

@article{apolinario_MeltingTransitions_2007,
  title = {Melting Transitions in Isotropically Confined Three-Dimensional Small {{Coulomb}} Clusters},
  author = {Apolinario, S. W. S. and Peeters, F. M.},
  year = 2007,
  month = sep,
  journal = {Physical Review E},
  volume = {76},
  number = {3},
  pages = {031107},
  issn = {1539-3755, 1550-2376},
  doi = {10.1103/PhysRevE.76.031107},
  urldate = {2026-03-27},
  copyright = {http://link.aps.org/licenses/aps-default-license},
  langid = {english}
}

@article{ludwig_StructureSpherical_2005,
  title = {Structure of Spherical Three-Dimensional {{Coulomb}} Crystals},
  author = {Ludwig, P. and Kosse, S. and Bonitz, M.},
  year = 2005,
  month = apr,
  journal = {Physical Review E},
  volume = {71},
  number = {4},
  pages = {046403},
  issn = {1539-3755, 1550-2376},
  doi = {10.1103/PhysRevE.71.046403},
  urldate = {2026-03-30},
  copyright = {http://link.aps.org/licenses/aps-default-license},
  langid = {english}
}

@article{apolinario_MultipleRings3D_2008,
  title = {Multiple Rings in a {{3D}} Anisotropic {{Wigner}} Crystal: {{Structural}} and Dynamical Properties},
  shorttitle = {Multiple Rings in a {{3D}} Anisotropic {{Wigner}} Crystal},
  author = {Apolinario, S. W. S. and Partoens, B. and Peeters, F. M.},
  year = 2008,
  month = jan,
  journal = {Physical Review B},
  volume = {77},
  number = {3},
  pages = {035321},
  issn = {1098-0121, 1550-235X},
  doi = {10.1103/PhysRevB.77.035321},
  urldate = {2026-03-30},
  copyright = {http://link.aps.org/licenses/aps-default-license},
  langid = {english}
}

@article{apolinario_StructuralDynamicalAspects_2007,
  title = {Structural and Dynamical Aspects of Small Three-Dimensional Spherical {{Coulomb}} Clusters},
  author = {Apolinario, S W S and Partoens, B and Peeters, F M},
  year = 2007,
  month = aug,
  journal = {New Journal of Physics},
  volume = {9},
  number = {8},
  pages = {283--283},
  issn = {1367-2630},
  doi = {10.1088/1367-2630/9/8/283},
  urldate = {2026-03-27}
}

@article{tsuruta_BindingEnergyMicrostructure_1993,
  title = {Binding Energy, Microstructure, and Shell Model of {{Coulomb}} Clusters},
  author = {Tsuruta, Kenji and Ichimaru, Setsuo},
  year = 1993,
  month = aug,
  journal = {Physical Review A},
  volume = {48},
  number = {2},
  pages = {1339--1344},
  issn = {1050-2947, 1094-1622},
  doi = {10.1103/PhysRevA.48.1339},
  urldate = {2026-04-08},
  copyright = {http://link.aps.org/licenses/aps-default-license},
  langid = {english}
}

@article{schweigert_SpectralPropertiesClassical_1995,
  title = {Spectral Properties of Classical Two-Dimensional Clusters},
  author = {Schweigert, Vitaly A. and Peeters, Fran{\c c}ois M.},
  year = 1995,
  month = mar,
  journal = {Physical Review B},
  volume = {51},
  number = {12},
  pages = {7700--7713},
  issn = {0163-1829, 1095-3795},
  doi = {10.1103/PhysRevB.51.7700},
  urldate = {2026-04-08},
  copyright = {http://link.aps.org/licenses/aps-default-license},
  langid = {english}
}

@article{schweigert_RadialFluctuationInducedStabilization_2000,
  title = {Radial-{{Fluctuation-Induced Stabilization}} of the {{Ordered State}} in {{Two-Dimensional Classical Clusters}}},
  author = {Schweigert, I. V. and Schweigert, V. A. and Peeters, F. M.},
  year = 2000,
  month = may,
  journal = {Physical Review Letters},
  volume = {84},
  number = {19},
  pages = {4381--4384},
  issn = {0031-9007, 1079-7114},
  doi = {10.1103/PhysRevLett.84.4381},
  urldate = {2026-04-08},
  copyright = {http://link.aps.org/licenses/aps-default-license},
  langid = {english}
}

@article{arp_3DCoulombBalls_2005,
  title = {{{3D Coulomb}} Balls: Experiment and Simulation},
  shorttitle = {{{3D Coulomb}} Balls},
  author = {Arp, O and Block, D and Bonitz, M and Fehske, H and Golubnychiy, V and Kosse, S and Ludwig, P and Melzer, A and Piel, A},
  year = 2005,
  month = jan,
  journal = {Journal of Physics: Conference Series},
  volume = {11},
  pages = {234--247},
  issn = {1742-6588, 1742-6596},
  doi = {10.1088/1742-6596/11/1/023},
  urldate = {2026-04-08}
}

@article{tomecka_MultistepRadialMelting_2005,
  title = {Multistep Radial Melting in Small Two-Dimensional Classical Clusters},
  author = {Tomecka, D. M. and Partoens, B. and Peeters, F. M.},
  year = 2005,
  month = jun,
  journal = {Physical Review E},
  volume = {71},
  number = {6},
  pages = {062401},
  issn = {1539-3755, 1550-2376},
  doi = {10.1103/PhysRevE.71.062401},
  urldate = {2026-04-08},
  copyright = {http://link.aps.org/licenses/aps-default-license},
  langid = {english}
}

@article{skeel_ImpulseIntegratorLangevin_2002,
  title = {An Impulse Integrator for {{Langevin}} Dynamics},
  author = {Skeel, Robert D. and Izaguirre, Jes{\"u}s A.},
  year = 2002,
  month = dec,
  journal = {Molecular Physics},
  volume = {100},
  number = {24},
  pages = {3885--3891},
  issn = {0026-8976, 1362-3028},
  doi = {10.1080/0026897021000018321},
  urldate = {2026-04-15},
  langid = {english}
}

@article{savitzky_SmoothingData_1964,
  title = {Smoothing and {{Differentiation}} of {{Data}} by {{Simplified Least Squares Procedures}}.},
  author = {Savitzky, {\relax Abraham}. and Golay, M. J. E.},
  year = 1964,
  month = jul,
  journal = {Analytical Chemistry},
  volume = {36},
  number = {8},
  pages = {1627--1639},
  publisher = {American Chemical Society},
  issn = {0003-2700},
  doi = {10.1021/ac60214a047}
}

@book{hollander_StatisticalMethods_2015,
  title = {Nonparametric {{Statistical Methods}}},
  author = {Hollander, Myles and A. Wolfe, Douglas and Chicken, Eric},
  year = 2015,
  month = jul,
  series = {Wiley {{Series}} in {{Probability}} and {{Statistics}}},
  edition = {1},
  publisher = {Wiley},
  doi = {10.1002/9781119196037},
  urldate = {2026-05-18},
  copyright = {http://doi.wiley.com/10.1002/tdm\_license\_1.1},
  isbn = {978-0-470-38737-5 978-1-119-19603-7},
  langid = {english}
}

@misc{guthrie_NISTHandbook_2020,
  title = {{{NIST}}/{{SEMATECH}} e-{{Handbook}} of {{Statistical Methods}} ({{NIST Handbook}} 151)},
  author = {Guthrie, William F.},
  year = 2020,
  publisher = {{National Institute of Standards and Technology}},
  doi = {10.18434/M32189},
  urldate = {2026-05-18},
  copyright = {License Information       for NIST data},
  langid = {english}
}
\appendix
\section{Simulated Annealing}
\begin{figure*}
    \includegraphics[width=\textwidth]{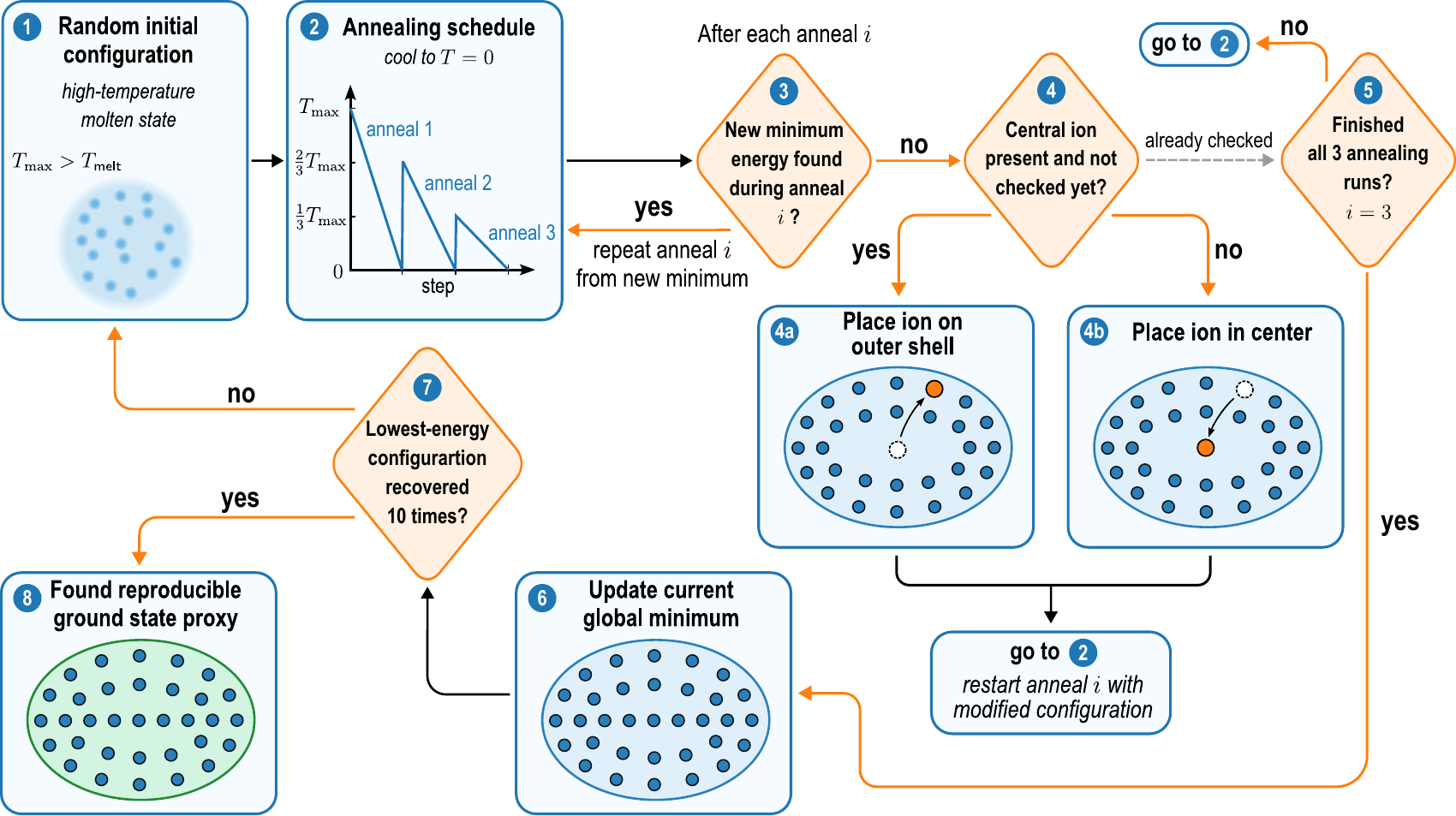}
    \caption{\justifying Schematic overview of the simulated-annealing procedure used to identify reproducible low-energy crystal configurations. Starting from a randomized high-temperature molten state (panel 1), the system is cooled to $T=0$ through a sequence of annealing schedules with successively reduced starting temperatures (panel 2). If a new minimum energy has been found during an anneal $i$, the anneal gets repeated from this new configuration (panel 3). Otherwise the minimum configuration is checked for a central ion to probe different shell occupancies (panel 4). If a central ion is present, it is displaced toward the outer shell (panel 4a). If no central ion is present, one ion is placed in the trap center (panel 4b). The current anneal $i$ is then restarted from the modified configuration. If this check has already been performed for the current anneal and all annealing stages are finished, the schedule is stopped accordingly (panel 5). Each candidate minimum is compared with the current global minimum which gets updated if necessary (panel 6). The search is stopped after the lowest-energy configuration has been recovered 10 times. Otherwise, the search continues from new randomized or modified initial configurations.
}
    \label{fig:simulated_annealing}
\end{figure*}
\begin{figure*}
    \includegraphics[width=0.75\textwidth]{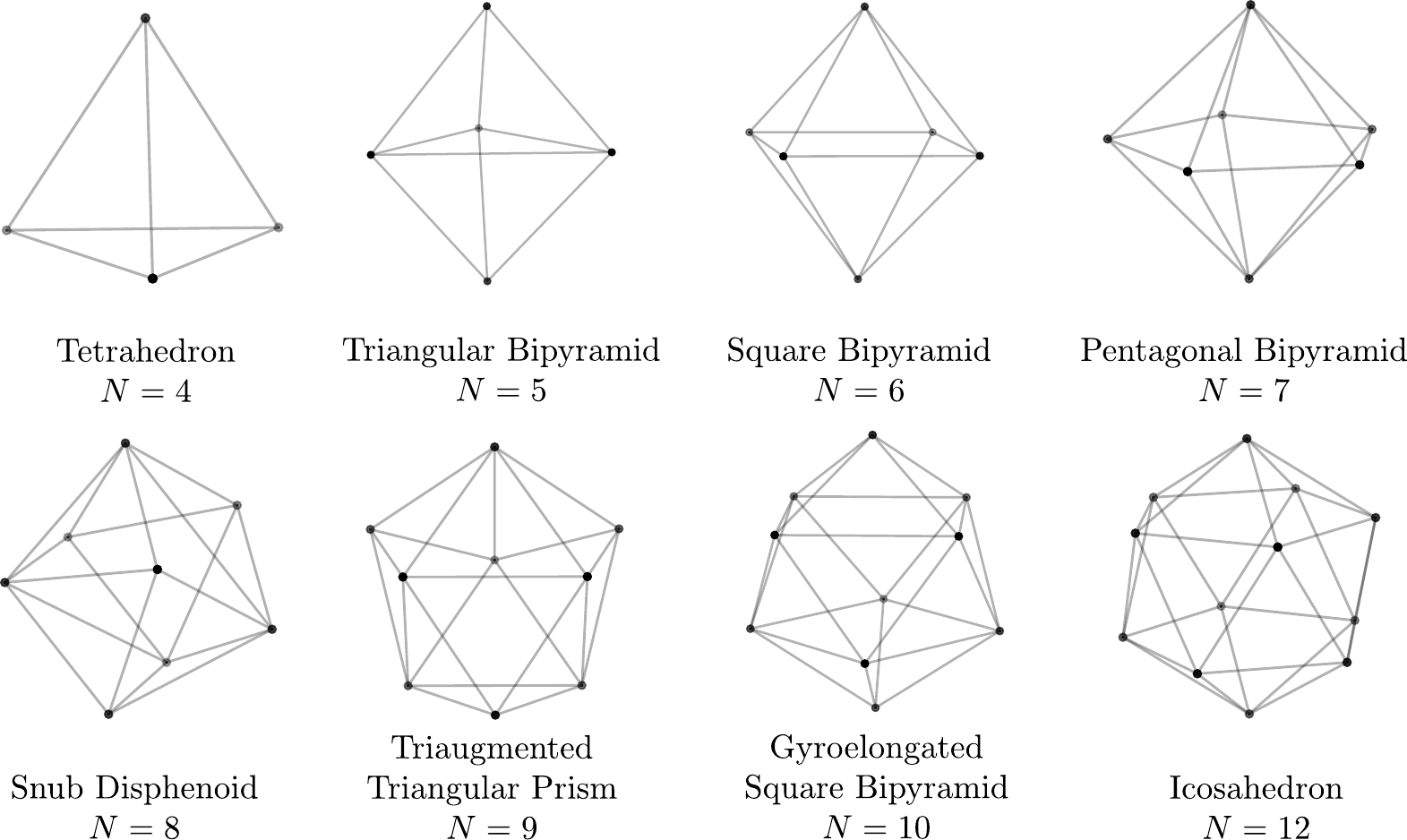}
    \caption{\justifying Deltahedron shapes based on the ground states of ion crystals with trapping parameter $\alpha=1$ for ion configurations up to $N=12$. Note that the $N=11$ configuration does not fit the definition of a deltahedron, which faces only consist of equilateral triangles. These shapes continue to appear in larger crystals when the innermost shells form.}
    \label{fig:deltahedrons}
\end{figure*}
\label{sec:ground state_prep}
To explore potential ground state configurations, we employ a simulated annealing (SA) algorithm \cite{kirkSimulatedAnnealing1983, caracciolo_SA_2023, solis_annealing_2014}, schematically illustrated in Fig.~\ref{fig:simulated_annealing}. Starting from a randomized configuration at a temperature where the crystal is in a molten state, the system is cooled stepwise to $T=0$. At each temperature step, the system is first equilibrated at the current temperature and then propagated further while sampling the potential energy and the corresponding configurations. From these samples, the lowest-energy configuration is stored, and the mean potential energy of the temperature step is compared to that of the previous temperature step. If the mean energy has decreased, the lowest-energy configuration sampled at the current temperature is accepted as the starting point for the next step. If the mean energy has increased, the configuration is accepted with the probability
\begin{equation}
    p_A = \min\left(1, \exp\left[- \frac{\Delta E}{k_B T}\right]\right),
\end{equation}
where $k_B$ is the Boltzmann constant, $T$ is the current temperature, and $\Delta E$ denotes the energy difference between successive temperature steps. This allows the algorithm to preferentially follow energetically favorable pathways while still permitting occasional transitions to higher-energy states, thereby exploring a wider energy landscape.

The annealing procedure of a complete ground state search cycle is repeated for at least three stages with successively reduced starting temperatures, while each stage is repeated if a new lower-energy state has been found. After each stage, the resulting candidate minimum is used to test different shell occupancies. If no ion is located close to the trap center, one ion is temporarily placed and fixed at the center before the annealing procedure is restarted. Conversely, if a central ion is present, this ion is displaced toward the outer shell before relaxation and subsequent annealing. These systematic modifications help to explore configurations with and without a central ion, reducing the risk of missing low-energy states with different shell occupations.

The complete ground state search cycle is repeated until the currently lowest-energy configuration has been recovered at least 10 times. Throughout this work, the term ground state therefore refers to the lowest-energy configuration found reproducibly across repeated SA runs from randomized and systematically altered initial conditions. While SA is a powerful method for exploring the energy landscape and reliably converges to low-energy configurations, it cannot guarantee that the true global minimum has been found. We therefore cannot exclude the existence of metastable configurations with slightly different energies or frictional properties. However, the reproducibility of the selected configurations across repeated searches provides a robust practical basis for the analysis presented here.

Geometric symmetries in the resulting crystal configurations can serve as an additional consistency check. For example, the inner shells of Coulomb crystals often forms characteristic deltahedral structures in low-energy configurations, as illustrated in Fig.~\ref{fig:deltahedrons}.

\section{Shell Analysis}
\label{sec:shell_analysis}
\begin{figure*}[t]
    \includegraphics[width=0.9\textwidth]{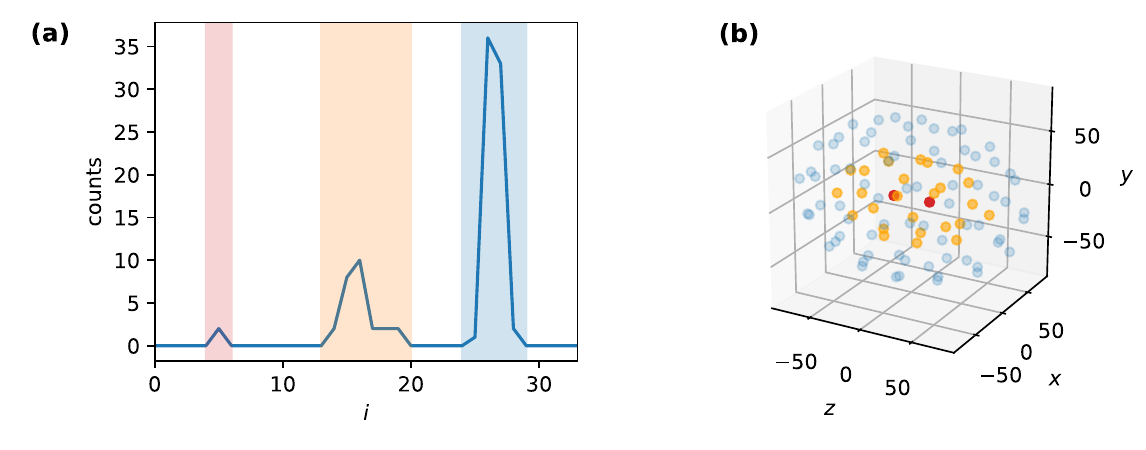}
    \caption{\justifying(a) Histogram of a 100-ion crystal with the iteration variable $i$ of the spheroidal shells (see eq.~\eqref{eq:shell_counting_1} and \eqref{eq:shell_counting_2}). The spacing $d$ was set to $\SI{2}{\micro\metre}$. The colored sections highlight the three shell regions. (b) 3D representation of the corresponding crystal. The color of the ions references the shell regions given in the histogram. The secular frequencies in the simulation are set to $\omega_{x,y,z}/2\pi=\{1.23, 1.23, 1\}\times \SI{180}{\kilo\hertz}$.}
    \label{fig:shell_structure_hist}
\end{figure*}
To characterize the shell structure of any given Coulomb crystal, we first fit a spheroid to the ion distribution. This fitting provides the semi-major axis $a$ and the semi-minor axis $b$. Subsequently, we iteratively define an inner and an outer spheroidal volume with semi-axes 
\begin{equation}
\label{eq:shell_counting_1}
L^\text{in}_i = (i+1) \cdot \frac{s_{ab}}{a} \cdot d
\end{equation}
for the outer ellipsoid and 
\begin{equation}
\label{eq:shell_counting_2}
L^\text{out}_i = i \cdot \frac{s_{ab}}{a} \cdot d 
\end{equation}
for the inner ellipsoid, respectively. Here, $i$ is an iteration variable that increases with each step, $s_{ab}$ represents the respective semi-major or minor axes and $d$ is the spacing between the two ellipsoidal shells. 

In each iteration, we count the number of ions located between the inner and outer spheroids and compile these counts into a histogram. This histogram reveals the shell structure of the crystal, providing insights into its spatial organization. This method is universally applicable to crystals with a spheroidal shape and ensures consistency across varying trap parameters. Fig.~\ref{fig:shell_structure_hist} shows an example of this shell analysis in the case of a crystal of $100$ ions.

\section{Inner-shell geometries}
\label{sec:inner_shell_structure}
\begin{figure*}
    \includegraphics[width=\textwidth]{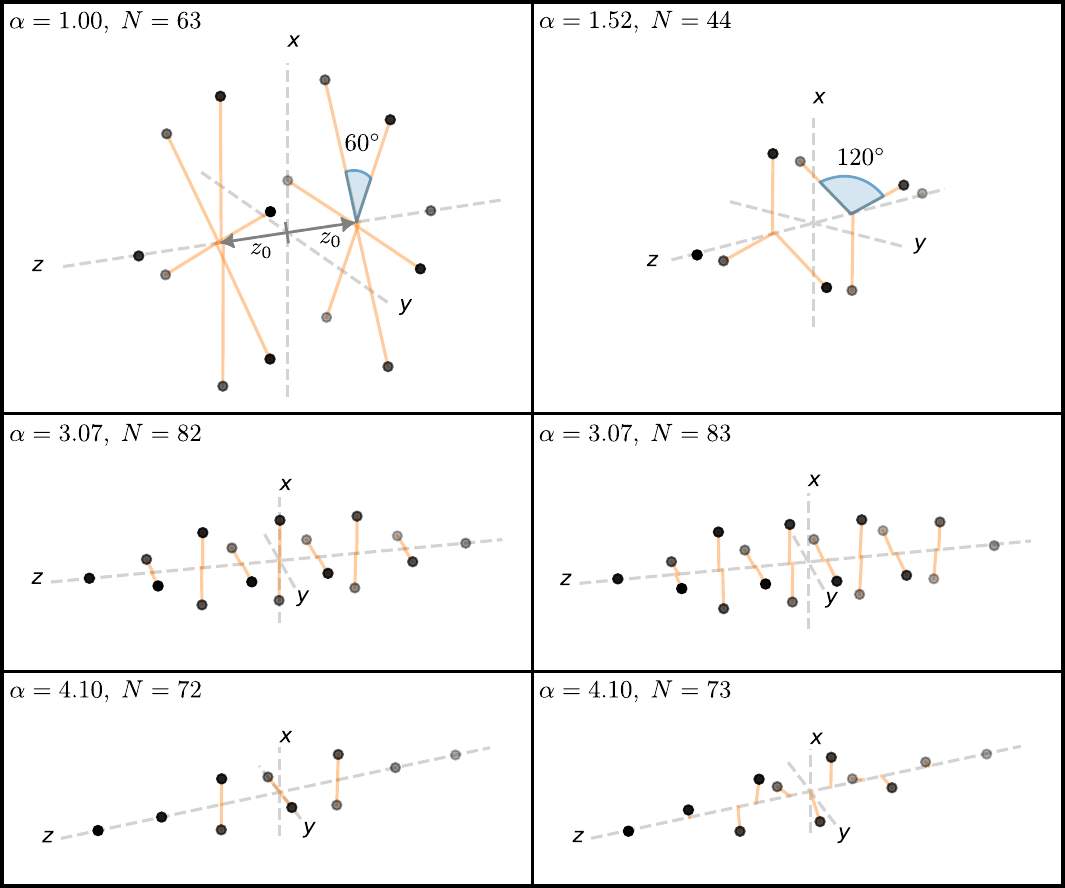}
    \caption{\justifying Configurations of the inner shell ions for selected crystals. Orange lines mark the projection of the ion positions onto the $z$-axis, while the opacity of the ions decreases with distance from the viewer to aid spatial visualization. In the $\alpha = 1.0$ case, the ions positions can be approximated by the vertices of an gyroelongated hexagonal bipyramid, which consists of two parallel hexagons in the $xy$-plane, equally spaced from the origin by $z_0$ and rotated by $30^\circ$ relative to each other. The exterior angle of $60^\circ$ is highlighted in blue. A similar arrangement is found in the $\alpha = 1.52$ system, where the ions approximate a gyroelongated triangular bipyramid with an exterior angle of $120^\circ$. For $\alpha = 3.07$ and $\alpha = 4.10$, the inner shell structure depends on the ion number and generally takes the form of double helical arrangements along the $z$-axis. Except for the $\alpha=4.10$, $N=73$ case, the double helical structures consist of pairs of ions, equidistant to the $z$-axis and alternating in their rotation around $z$ in steps of $90^\circ$. For the $\alpha=4.10$, $N=73$ system, this symmetry is broken due to an additional ion in the inner shell.}
    \label{fig:inner_shell_configs}
\end{figure*}

To provide a structural reference for the systems discussed in Sec.~\ref{subsec:shell_rotation}, we show the corresponding inner-shell configurations in Fig.~\ref{fig:inner_shell_configs}.

For $\alpha = 1.0$ with $N = 63$ and $N = 64$, as well as for $\alpha = 1.52$ with $N = 44$ and $N = 50$, the inner shells are nearly identical, differing only by slight positional shifts. We therefore display only one representative configuration for each pair.

In general, ions confined by an isotropic trapping potential tend to form structures from nearly equilateral triangles. For the two $\alpha = 1.0$ systems, this leads to a geometry that can be approximated by a gyroelongated hexagonal bipyramid, while the $\alpha = 1.52$ systems are well described by a gyroelongated triangular bipyramid. The characteristic exterior angles of these structures are reflected in the periodicities of the relaxed total energy $E(\varphi)$ shown in Fig.~\ref{fig:E_pot_vs_theta}, illustrating how the inner-shell geometry can imprint a corresponding corrugation pattern onto the rotational path of the outer shell.

For the $\alpha = 3.07$ systems, the inner shell forms a double helix around the $z$-axis, with ion pairs that are equidistant from the $z$-axis and alternate in $90^\circ$ steps in the $xy$-plane. Notably, adding a single ion increases the inner-shell population by two.

In the $\alpha = 4.10$, $N = 72$ system, the inner shell also exhibits a helical structure that flattens into linear chains towards the edges. Adding one ion increases the inner-shell count by one, causing a splitting of the alternating ion pairs along $z$ and variations in their orientation within the $xy$-plane.

These examples illustrate that the rotational energy landscape can be modified by structural changes of the inner shell, the outer shell, or both. In the $\alpha = 3.07$ and $\alpha = 4.10$ systems, the inner shells reconfigure with changing ion number, whereas for $\alpha = 1.0$ and $\alpha = 1.52$ the inner shell remains nearly unchanged and the main structural differences occur in the outer shell. While the inner shell geometry has a significant impact on the Peierls-Nabarro-type energy landscape of the outer shell rotation, the resulting effective barrier is a composite quantity that reflects the combined inter-shell, outer-shell, and trap-related energy changes along the relaxed rotational path, as discussed in Appendix~\ref{sec:dynamic_commensurability}.

\section{Comparison of second energy difference and effective energy barrier}
\label{sec:second_energy_comparison}
\begin{figure*}[t]
    \centering
    \includegraphics[width=\textwidth]{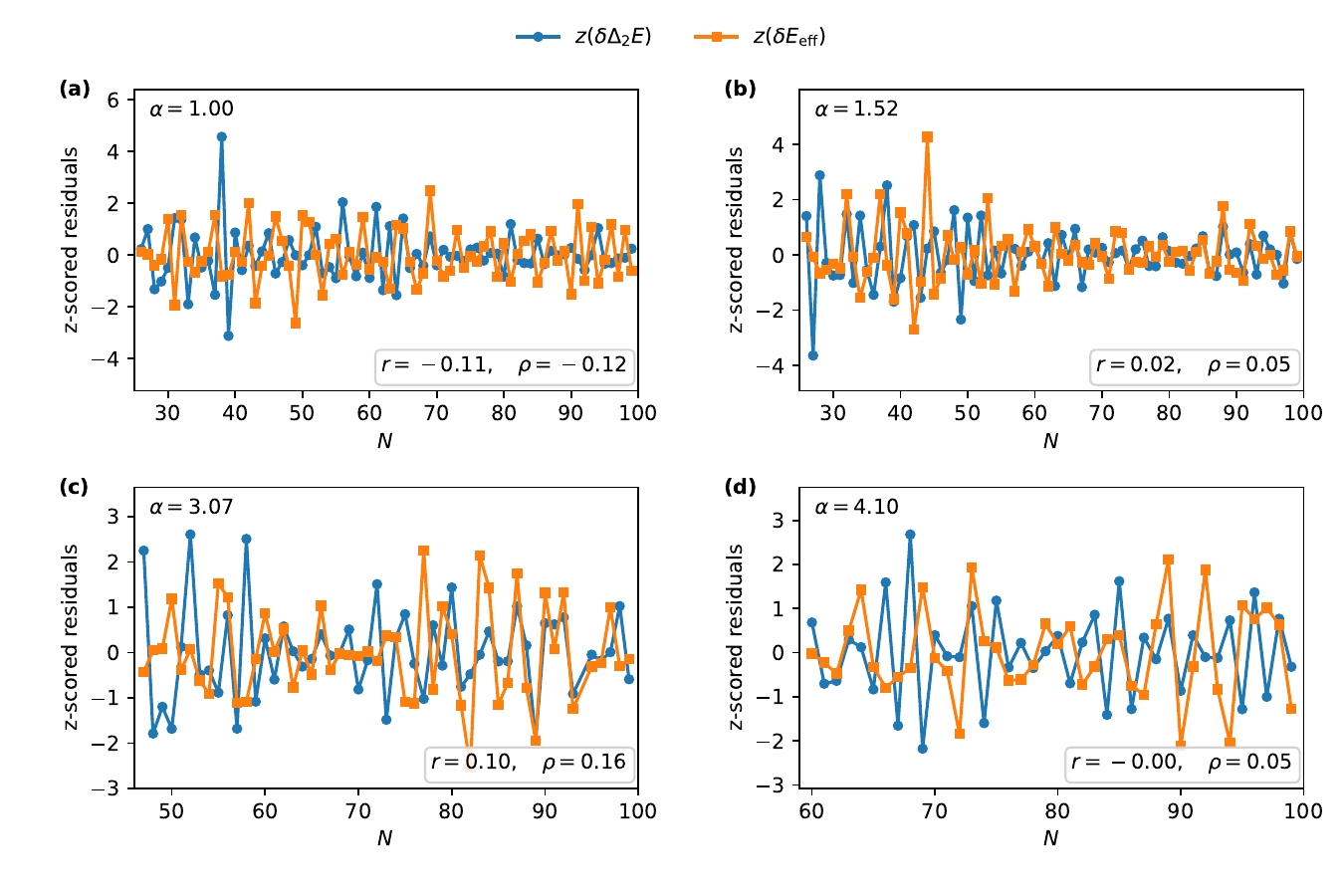}
    \caption{
    Comparison of locally detrended fluctuations of the second energy difference $\Delta_2E(N)$ and the normalized effective barrier $E_{\mathrm{eff}}(N)$ for different trap aspect ratios. For each $\alpha$, both series are detrended, and converted to z-scored residuals. The plotted curves therefore show only the local deviations from the smooth background trend as a function of $N$. The Pearson coefficient $r$ and the Spearman rank coefficient $\rho$, shown in each panel, remain small in magnitude for all four systems, confirming that the local fluctuations of $\Delta_2E(N)$ do not exhibit a systematic correlation with those of $E_{\mathrm{eff}}(N)$.}
    \label{fig:second_energy_difference}
\end{figure*}
To test whether local variations of the effective rotational barrier are related to the finite-size stability of the corresponding ground states, we compare the second energy difference
\begin{equation}
\Delta_2E(N)=E_{\mathrm{tot}}(N+1)+E_{\mathrm{tot}}(N-1)-2E_{\mathrm{tot}}(N)
\end{equation}
to the normalized effective barrier $E_{\mathrm{eff}}(N)$ for the same values of $\alpha$ and over the same $N$ ranges as used in Fig.~\ref{fig:eff_barrier}.

Rather than comparing the raw series directly, we focus on their local fluctuations. For each series $y(N)$, we define the residual
\begin{equation}
\delta y(N)=y(N)-y_{\mathrm{bg}}(N),
\end{equation}
where $y_{\mathrm{bg}}(N)$ is a smooth background obtained from a local Savitzky-Golay filter. This filter performs a local polynomial smoothing by fitting a low-order polynomial within a moving window around each data point~\cite{savitzky_SmoothingData_1964}.
We then calculate the z-scored residuals to bring both series onto a common dimensionless scale,
\begin{equation}
z_{\delta y}(N)=\frac{\delta y(N)-\langle \delta y \rangle}{\sigma_{\delta y}},
\end{equation}
where $\langle \delta y \rangle$ and $\sigma_{\delta y}$ denote the mean and standard deviation of the residual series. Thus, $z_{\delta y}(N)$ measures the local deviation from the smooth background in units of the standard deviation of the residual fluctuations. This normalization allows for a direct comparison of the fluctuations of $\Delta_2E(N)$ and $E_{\mathrm{eff}}(N)$ despite their different scales.

The z-scored residuals of both series are shown in Fig.~\ref{fig:second_energy_difference}. If the pronounced barrier variations were primarily controlled by magic-number stability, one would expect peaks and dips of both residual series to coincide. Instead, the maxima and minima generally do not align.

To quantify the correlation between both residual series, we compute the Pearson correlation coefficient $r$ and the Spearman rank correlation coefficient $\rho$ for each $\alpha$. The Pearson coefficient is defined as
\begin{equation}
r =
\frac{\sum_i (x_i-\langle x\rangle)(y_i-\langle y\rangle)}
{\sqrt{\sum_i (x_i-\langle x\rangle)^2}\sqrt{\sum_i (y_i-\langle y\rangle)^2}},
\end{equation}
where $x_i=z_{\delta \Delta_2E}(N_i)$ and $y_i=z_{\delta E_{\mathrm{eff}}}(N_i)$ are the matched residual values at common particle numbers $N_i$. It measures the linear correlation between both residual series, with $r=1$ corresponding to perfect positive linear correlation, $r=-1$ to perfect negative linear correlation, and $r=0$ to no linear correlation~\cite{guthrie_NISTHandbook_2020}. 

The Spearman coefficient $\rho$ is defined as the Pearson correlation coefficient of the rank variables $R(x_i)$ and $R(y_i)$, where $R(x_i)$ assigns to each value $x_i$ its position in the sorted set of all $x$ values~\cite{hollander_StatisticalMethods_2015}. It therefore measures whether both series are monotonically related, independent of the precise linear scaling of their values.

As shown in Fig.~\ref{fig:second_energy_difference}, both coefficients remain small in magnitude for all four cases, indicating that local fluctuations in $\Delta_2E(N)$ do not systematically explain the local fluctuations of $E_{\mathrm{eff}}(N)$.

\section{Dynamic inter-shell commensurability}
\label{sec:dynamic_commensurability}
\begin{figure*}
    \includegraphics[width=0.9\textwidth]{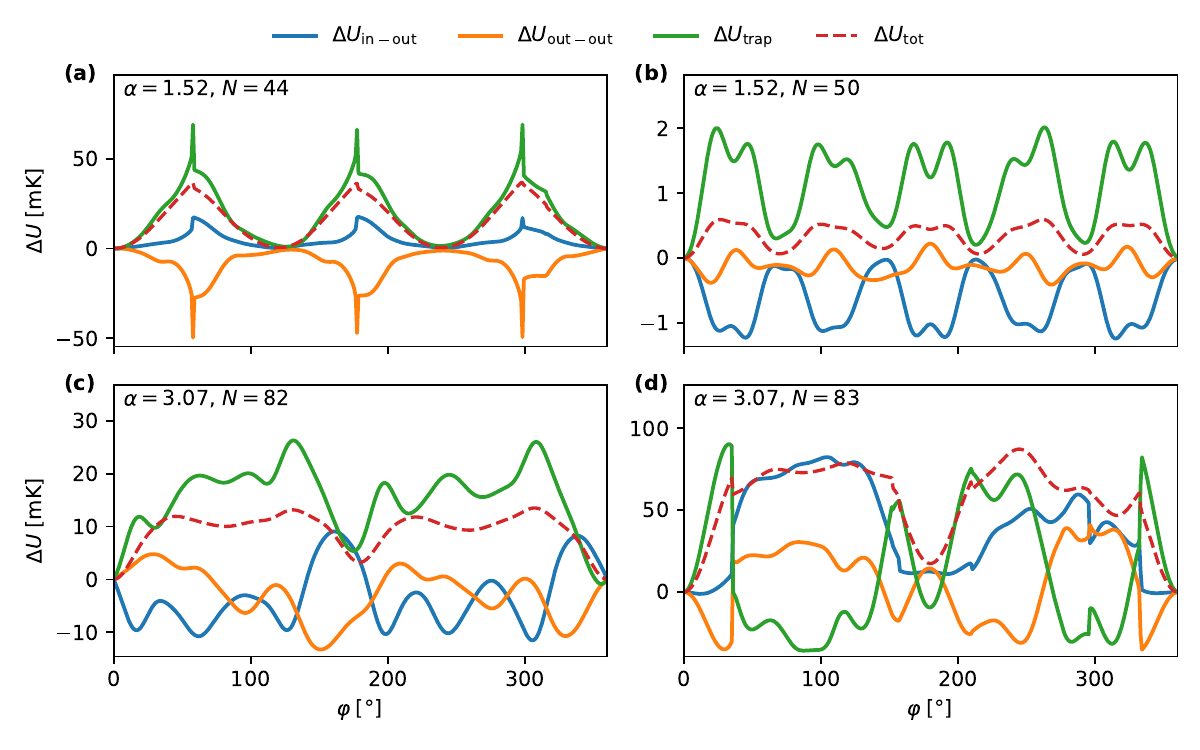}
    \caption{\justifying Decomposition of the effective energy barrier along the relaxed rotational path for two representative pairs of systems discussed in the main text: $\alpha=1.52$ with $N=44$ and $N=50$, and $\alpha=3.07$ with $N=82$ and $N=83$. Shown are the energy changes $\Delta U_k(\varphi)=U_k(\varphi)-U_k(0)$ as a function of the rotation angle $\varphi$ of the driven ions on the outer shell. $\Delta U_{\text{in-out}}$ (blue) denotes the interaction energy change between inner and outer shell, $\Delta U_{\text{out-out}}$ (orange) the interaction energy change within the outer shell, and $\Delta U_{\mathrm{trap}}$ (green) the change in trap energy. The red dashed line shows the corresponding total energy change, $\Delta U_{\mathrm{tot}}(\varphi)$. The examples illustrate that the effective barrier is a composite quantity: depending on the system, its variation can be dominated either by the inter-shell interaction or by the trap-related cost of the outer-shell relaxation, while the individual terms may also partially compensate each other.}
    \label{fig:energy_decomposition}
\end{figure*}

\begin{figure*}
    \includegraphics[width=0.9\textwidth]{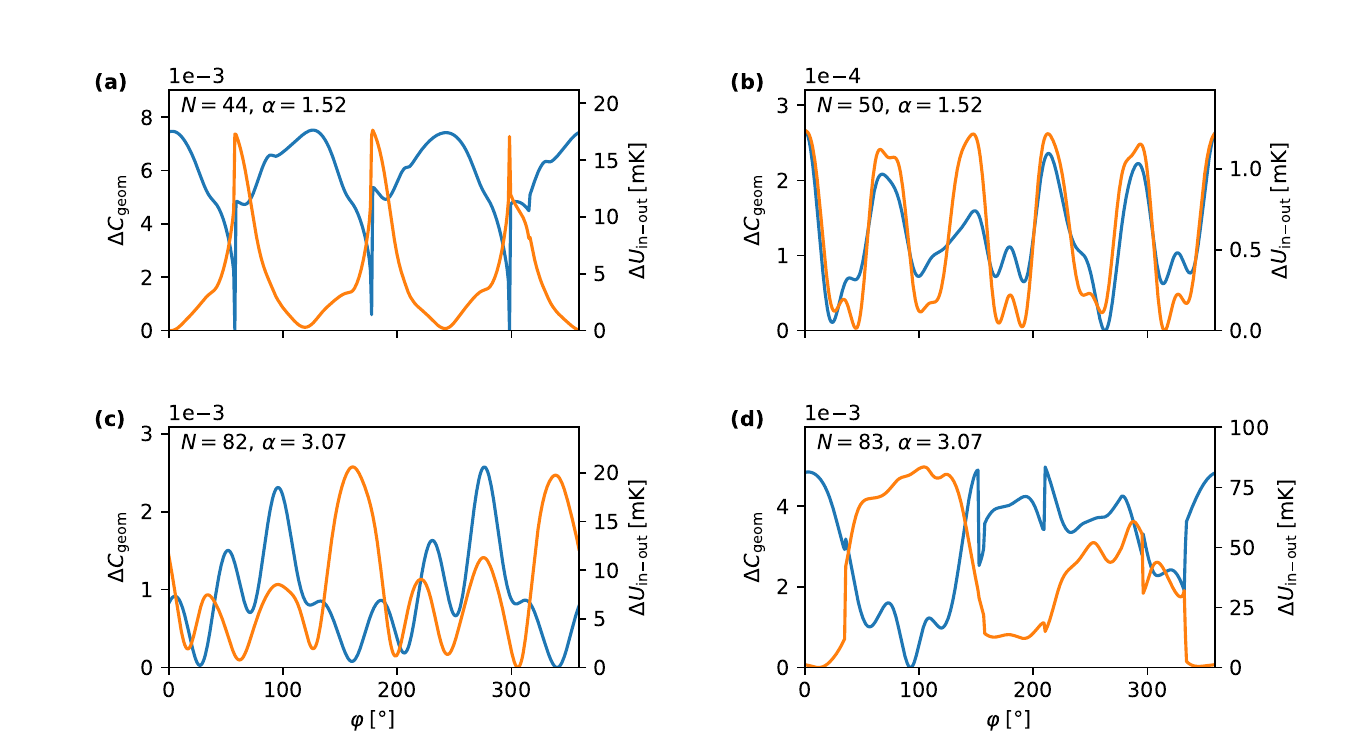}
    \caption{\justifying
    Comparison of the geometric commensurability modulation and the inter-shell interaction corrugation along the relaxed rotational path for four representative systems. Shown are the shifted quantities ${\Delta C_{\mathrm{geom}}(\varphi)=C_{\mathrm{geom}}(\varphi)-\min (C_{\mathrm{geom}})}$ (blue) and ${\Delta U_{\text{in-out}}(\varphi)=U_{\text{in-out}}(\varphi)-\min (U_{\text{in-out}})}$ (orange) as functions of the rotation angle $\varphi$ of the driven ions on the outer shell. Panels (a) and (b) correspond to $\alpha=1.52$ with $N=44$ and $N=50$, respectively, while panels (c) and (d) show $\alpha=3.07$ with $N=82$ and $N=83$. The geometric commensurability factor $C_{\mathrm{geom}}(\varphi)$ is evaluated from the void structure of the inner shell on a common spheroidal reference surface and quantifies how strongly the outer-shell ions occupy geometrically favorable interstitial regions. In panels (a) and (d), minima of $\Delta C_{\mathrm{geom}}(\varphi)$ coincide with maxima of $\Delta U_{\text{in-out}}(\varphi)$, indicating that geometrically unfavorable commensurability enhances the inter-shell interaction energy. In panels (b) and (c), this correspondence is weaker, showing that the projected geometric commensurability captures an important, but not always exclusive, contribution to the variation of $U_{\text{in-out}}(\varphi)$ along the relaxed path.
    }
    \label{fig:dynamic_commensurability_selected}
\end{figure*}

To better understand the pronounced fluctuations in the effective barrier $E_\text{eff}$ discussed in Sec.~\ref{subsec:energy_barrier}, it is useful to separate the different contributions that build up the barrier along the relaxed rotational path. The effective barrier does not only reflect the mutual commensurability of two rigid shells. It contains contributions from the interaction between the shells, the mutual Coulomb interaction of the ions in the outer shell which are allowed to relax after each rotation step, and the energy of the ions in the confining trapping potential.

Figure~\ref{fig:energy_decomposition} shows the decomposition of the total potential energy variations along the relaxed path: The inter-shell contribution $\Delta U_{\text{in-out}}$, the intra-shell contribution of the outer shell $\Delta U_{\text{out-out}}$, and the trap contribution $\Delta U_{\mathrm{trap}}$. This decomposition allows us to directly assess to what extent the changes in $E_\text{eff}$ with ion number can be attributed to a change in inter-shell commensurability, and to what extent it originates from the relaxation of the outer shell in the confining potential. The examples show that the relative importance of these terms is strongly system dependent. In some systems the barrier is strongly influenced by the inter-shell interaction (such as in the $N=83$, $\alpha=3.07$ case), whereas in others the trap-related cost of the structural relaxation of the outer shell provides the dominant contribution. The effective barrier $E_\text{eff}$ is therefore a composite quantity and cannot be interpreted as a direct measure of inter-shell commensurability alone.

To quantify the geometric inter-shell commensurability independently of these energetic contributions, we introduce a dynamic geometric commensurability factor $C_{\mathrm{geom}}(\varphi)$ that is evaluated along the same relaxed rotational path. For each relaxed configuration, the ion positions of both shells are projected onto a common spheroidal reference surface, which provides a two-dimensional parametrization of the shell geometry.

On this reference surface, we can define a continuous occupancy field of the ions of the inner shell
\begin{equation}
O_{\mathrm{in}}(s)=\sum_{i \in \mathrm{in}}
\exp\!\left[-\frac{d_g(s,s_i)^2}{2\sigma^2}\right],
\end{equation}
where $s_i$ denotes the projected position of the $i$-th ion on the inner shell and $d_g$ is the geodesic distance on the spheroidal surface. The length scale $\sigma$ is chosen as $\sigma = 0.35\, d_{\mathrm{nn,in}}$, where $d_{\mathrm{nn,in}}$ is the mean geodesic nearest-neighbor spacing of the inner shell, such that the void structure between neighboring ions remains spatially resolved while still providing a smooth continuous field.
From this occupancy field we construct the corresponding void field
\begin{equation}
G_{\mathrm{in}}(s)=1-\frac{O_\mathrm{in}(s)}{\max (O_\mathrm{in})}.
\end{equation}
High values of $G_{\mathrm{in}}(s)$ identify void regions between inner-shell ions, whereas low values correspond to coordinates directly on top of the inner-shell particle pattern. The geometric commensurability factor $C_\text{geom} (\varphi)$ as a function of the angle of outer-shell rotation $\varphi$ is then defined as the mean value of this void field at the projected positions $s_j^{\mathrm{out}}(\varphi)$ of the outer-shell ions,
\begin{equation}
C_{\mathrm{geom}}(\varphi)=
\frac{1}{N_{\mathrm{out}}}
\sum_{j \in \mathrm{out}}
G_{\mathrm{in}}\!\left(s_j(\varphi)\right).
\end{equation}
In this way, commensurability is quantified as the degree to which the outer shell occupies the void structure of the inner shell with $C_\text{geom} (\varphi)$ yielding a continuous structural correlation measure of the evolving shell commensurability. A large value of $C_\text{geom}$ corresponds to a high commensurability between the shells with ions of the outer shell preferentially occupying the void structure between ions of the inner shell, and vice versa. 

Fig.~\ref{fig:dynamic_commensurability_selected} compares the geometric commensurability factor $C_\text{geom}$ with the inter-shell interaction corrugation $U_\text{in-out}$ as functions of the rotation angle $\varphi$ for selected systems. In panels (a) and (d), minima of $\Delta C_{\mathrm{geom}}(\varphi)$ coincide with maxima of $\Delta U_{\mathrm{in-out}}(\varphi)$, indicating that geometrically unfavorable commensurability enhances the inter-shell interaction energy. In panels (b) and (c), this correspondence is weaker, showing that the projected geometric commensurability captures an important, but not always exclusive, contribution to the variation of $U_{\mathrm{in-out}}(\varphi)$ along the relaxed path. While $C_{\mathrm{geom}}(\varphi)$ quantifies the structural matching of the two shells on a common spheroidal reference surface, $U_{\mathrm{in-out}}(\varphi)$ is determined by the full three-dimensional Coulomb interaction and can therefore additionally depend on subtle relaxations of the outer shell that are only weakly reflected in the projected geometric measure.

Taken together, the two analyses provide complementary information. The energy decomposition identifies which energetic contributions dominate the full barrier in a given system, while the geometric commensurability factor isolates how strongly the direct structural matching between inner and outer shell varies along the sliding coordinate. The comparison shows that variations in inter-shell commensurability can account for an important part of the variation of $\Delta U_{\text{in-out}}(\varphi)$ in some systems, but that this correspondence is not universal. While the geometric commensurability captures a relevant structural contribution to the inter-shell interaction corrugation, the detailed behavior of $\Delta U_{\text{in-out}}(\varphi)$ can additionally be influenced by the full three-dimensional arrangement of the ions and by subtle relaxations of the outer shell. The full effective barrier $E_\text{eff}$ furthermore depends on the structural response of the outer shell in the trap and is therefore not determined by geometric commensurability alone.

\section{Discussion of depinning–torque ratios}
\label{sec:depinning_torque}
\begin{table}
\centering
\begin{tabular}{c @{\hspace{1em}} c @{\hspace{1em}} c @{\hspace{1em}} c}
\hline
$\alpha$ & $N, N'$ & $\tau^N_{\mathrm{qs}}/\tau^{N'}_{\mathrm{qs}}$ & $\tau^N_{\mathrm{dyn}}/\tau^{N'}_{\mathrm{dyn}}$ \\
\hline
1.00 & 64, 63 & $3.75$ & $7.72$ \\
1.52 & 50, 44 & $26.74$ & $3.00$ \\
3.07 & 82, 83 & $6.01$ & $7.24$ \\
4.10 & 72, 73 & $4.10$ & $3.15$ \\
\hline
\end{tabular}
\caption{\justifying Comparison of depinning–torque ratios for matched system pairs at fixed $\alpha$. For each pair $(N, N')$, the ratio of the depinning torques of the quasistatic (qs) and in the dynamical (dyn) case is given (ratios are given as larger threshold divided by smaller threshold for each pair). The quasistatic estimate is obtained from the maximum continuous slope of the Peierls-Nabarro-type $\Delta E(\varphi)$ calculated in Sec.~\ref{subsec:energy_barrier} where the driving–ion chain is constrained in the azimuthal direction. The dynamic threshold (dyn) is extracted from unconstrained Langevin simulations of the outer shell with an applied rotational torque.}
\label{tab:torque_ratios}
\end{table}
From the relaxed Peierls--Nabarro-type energy landscape $\Delta E(\phi)$ obtained in Sec.~IVB, we estimate a quasistatic torque scale from the maximum continuous slope,
\begin{equation}
\tau_{\mathrm{qs}}^{N}
=
\max_{\phi\in C}
\left|
\frac{\partial \Delta E^{N}(\phi)}{\partial \phi}
\right| ,
\end{equation}
where the set $C$ excludes angles at which reordering events produce non-differentiable changes in $\Delta E(\phi)$. This quantity should be interpreted as the generalized torque required to follow the constrained, relaxed rotational path used in the quasistatic calculation. It is therefore not identical to the inter-shell corrugation torque in the dynamical simulations. Since $\Delta E(\phi)$ is the fully relaxed total energy, $\tau_{\mathrm{qs}}$ also contains the energetic cost associated with structural relaxation of the outer shell and its motion in the confining trap potential.

In Sec.~IVC, the dynamical depinning threshold $\tau_{\mathrm{dyn}}$ is extracted from unconstrained Langevin simulations in which all outer-shell ions are driven by an applied rotational torque. The comparison between $\tau_{\mathrm{qs}}$ and $\tau_{\mathrm{dyn}}$ is therefore not expected to be quantitatively exact. The quasistatic calculation follows a prescribed rotational path with an azimuthally constrained driving-ion chain, whereas the dynamical system can exploit additional degrees of freedom, local rearrangements and deformations of the outer shell. As a result, the dynamically rotating shell may avoid the steepest sections of the quasistatic path or depin through a different trajectory in configuration space.

A concise comparison of the resulting torque ratios for matched system pairs at fixed $\alpha$ is given in Tab.~\ref{tab:torque_ratios}. Although the quasistatic and dynamical ratios differ quantitatively, they correctly reproduce the relative ordering of the selected systems: configurations with a larger quasistatic torque scale also exhibit larger dynamical depinning thresholds. The remaining differences reflect the fact that the quasistatic PN landscape provides an effective, path-dependent barrier measure, while depinning in the full Langevin dynamics is a high-dimensional process influenced by intra-shell coupling, damping, local rearrangements and, in some cases, metastable dynamical states.
\end{document}